\documentclass[accepted=2024-09-13
,a4paper,onecolumn,11pt]{quantumarticle}

\pdfoutput=1
\usepackage[dvipsnames]{xcolor}
\usepackage[utf8]{inputenc}
\usepackage[english]{babel}
\usepackage[T1]{fontenc}
\usepackage{amsmath}
\usepackage{hyperref}
\newcommand{\trr}{\textcolor{Black}}
\newcommand{\trs}{\textcolor{black}}

\usepackage{tikz}
\usepackage{lipsum}

\usepackage[numbers,sort&compress]{natbib}

\begin{document}

\title{Mitigating Fast Controller Noise in Quantum Gates Using Optimal Control Theory}

\author{Aviv Aroch}
\email{aviv.aroch@mail.huji.ac.il}
\affiliation{ 
The Institute of Chemistry, The Hebrew University of Jerusalem, Jerusalem 9190401, Israel}
\orcid{0000-0002-0508-1992}

\author{Shimshon Kallush}
\email{shimshonk@hit.ac.il}
\affiliation{Sciences Department, Holon Academic Institute of Technology, 52 Golomb Street, Holon 58102, Israel}
\orcid{0000-0002-9036-5964}

\author{Ronnie Kosloff}
\email{kosloff1948@gmail.com}
\affiliation{ 
The Institute of Chemistry, The Hebrew University of Jerusalem, Jerusalem 9190401, Israel}
\orcid{0000-0001-6201-2523}

\begin{abstract}
All quantum systems are subject to noise from the environment
or external controls. This noise is a major obstacle to the realization of quantum technology; e.g., noise 
limits the fidelity of quantum gates.
We employ Optimal Control Theory to study the generation of quantum single- and two-qubit gates in the presence of fast controller noise. Specifically, we explore a Markovian model of phase and amplitude noise, which degrades gate fidelity.
We show that optimal control with such noise models generates control solutions to mitigate the loss of gate fidelity. The problem is formulated in Liouville space, employing a highly accurate numerical solver and the Krotov algorithm to solve optimal control equations.
\end{abstract} 

\label{sec:intro}
In reality, any quantum system is open. External intervention transforms the unitary dynamics of a closed system into a non-unitary evolution of an open one.
We can classify three significant sources of such intervention:
I. A thermal environment.
II. Back-action due to quantum measurement.
III. Noise originating from the external controller.
The loss of coherence associated with quantum systems is the biggest challenge for achieving quantum technology \cite{preskill2018quantum,schlosshauer2019quantum}.
This study explores Optimal Control Theory (OCT) to mitigate the fidelity destruction of a quantum gate due to controller noise.

Quantum phenomena with an advantage in information processing are fragile. 
Combating errors due to hardware noise has, therefore, been a significant challenge. 
A software solution suggested is error correction, embedding the computation base in a much larger Hilbert space.  
The analysis has led to the threshold law \cite{aharonov1997fault}, 
defining the minimum fidelity required in the quantum gate to allow for error correction. 
This, in turn, poses a challenge to the hardware: a requirement of extremely high fidelity in the elementary gate operations.

The efforts to obtain high-fidelity quantum gates
can be roughly divided into passive and active strategies.
Passive proposals are based on finding a quiet region in Hilbert space, such as a decoherence-free subspace \cite{lidar1998decoherence} or topologically protected
states \cite{hasan2010colloquium}. Consequently, the computation tasks are performed by protocols that are relatively immune to noise. Gates based on  adiabatic dynamics \cite{lacour2008optimized,chen2012engineering,benseny2021adiabatic,genov2014correction}, invariant based \cite{ruschhaupt2012optimally,lu2014fast,levy2018noise,colmenar2022reverse} or inertial protocols \cite{turyansky2024inertial}
have been developed. 
Active error mitigation is based on noise analysis and active intervention to combat it. In typical cases, a time-scale separation 
can be identified between a slow, varying noise and a fast control. 
An active control solution
dresses the dynamical spectrum of the qubit system with a high-frequency modulation, effectively moving it out of the noise spectrum.
A scheme originating in nuclear
magnetic resonance, dynamical decoupling,
implements this idea for quantum gates
\cite{uhrig2007keeping,khodjasteh2009dynamically,viola1999dynamical,pokharel2018demonstration}.
Time-scale separation has also been a source for additional schemes 
to combat noise, such as combining noise filtering  
with geometric control \cite{li2021designing} or employing optimal control theory \cite{gutmann2005compensation,montangero2007robust,kabytayev2014robustness}. 

\trs{Another strategy to actively combat noise relies on a
Magnus or Dyson expansion of the noise propagator  \cite{green2013arbitrary,dong2023resource,shao2024multiple,auza2024quantum}. This scheme is designed to combat colored noise by canceling specific orders of the expansion based on knowledge of the noise correlation functions.}

\trs{The Markovian limit sets in when 
the timescale separation between the system and the noisy source is reversed. Such fast noise is the worst-case scenario. 
Can such noise be tamed? 
Optimal control theory has been employed to combat such noise \cite{kallush2006quantum,montangero2007robust,katz2007decoherence,shao2024multiple} with achievable results.
Recently, optimal control based on machine learning has been explored \cite{youssry2020characterization,zeng2020quantum}.
The mechanism of success remains an open question.
Some studies have revealed a control mechanism based on active cooling, which removes the noise to an entropy sink
\cite{kallush2006quantum,katz2007decoherence,kallush2022controlling}. }

\trs{Active control strategies to combat noise
come with a price, a new noise source 
originating from the controller.
Typically, the controller is a pulse generator responsible for
executing the quantum gates.
Scaling arguments suggest that
controller noise leads to purity loss that scales unfavorably 
with system size \cite{khasin2011noise,kallush2014quantum}. Can we combat the additional noise introduced by the controller?
For state-to-state transformations, optimized invariant-based noise mitigation has been suggested \cite{ruschhaupt2012optimally,levy2018noise}.}

\trs{For quantum information processing
the full power of optimal control theory has not been applied, with an exception addressing slow noise sources \cite{kabytayev2014robustness}. 
In the present study, we address the most challenging
problem of mitigating Markovian noise
originating from the controller in quantum gate execution. 
Analysis shows that this noise always contains a fast component \cite{haffner2008quantu,day2022limits,PhysRevA.108.012620,jiang2023sensitivity,kaushal2020shuttling}. Can we control the controller in such a way that minimizes this harmful effect?}

To address these issues, we formulate the dynamics of open quantum systems by considering a completely positive trace-preserving map (CPTP) \cite{kraus1971general}.
In the Markovian limit, CPTP maps are generated by the equation:
$$\frac{\partial}{\partial t} \hat{\boldsymbol{\rho}} ={\cal L} \hat{\boldsymbol{\rho}}$$
where the generator ${\cal L}$ has the
Gorini,
Kossakowski, Lindblad, and Sudarshan (GKLS) structure \cite{gorini1976n,lindblad1976generators}. We adopt this framework for the present study.

OCT has previously addressed open-system control problems 
\cite{Bartana-1997,ohtsuki1999monotonically,bartana2001laser,schulte2011optimal,koch2016controlling,Basilewitsch_2018,pokharel2018demonstration,petruhanov2023grape,PhysRevA.108.012620,petruhanov2023quantum,fernandes2023effectiveness}. In such control tasks, the dynamics generator ${\cal L}$ becomes explicitly time-dependent, i.e., a function of the control field ${\cal L}(\epsilon(t))$ \cite{huang2023completely}. This observation raises the issue: Are the dissipative and unitary dynamics linked? Does a change in the control field modify the dissipative dynamics?

Consistency with thermodynamics imposes additional restrictions on the form of the thermal environment's master equation \cite{dann2021quantum}.
In this case, the dissipative and the unitary parts are linked. Optimal control of quantum state-to-state and quantum gates were studied, employing a thermodynamically consistent master equation   \cite{kallush2022controlling}. It was demonstrated that both entropy-changing and preserving transformations can be achieved. In particular, the negative aspect of dissipation on the fidelity of quantum gates could be mitigated. The mechanism
identified could be described
as active refrigeration
accompanying the control process leading to a unitary quantum gate.

Our primary concern is to explore quantum control strategies that mitigate noise and allow the maintenance of high-fidelity quantum gates.
We first address this issue by employing OCT to generate unitary quantum gates in a pure Hamiltonian setup.  Both single and two-qubit gates are addressed. We choose gates that form a universal computation set. We concentrate on high-fidelity unitary gates, which we can obtain by our noiseless control.
We subsequently add the noise to the controller and observe the degradation of fidelity.
On this reference gate, we apply open system control, including noise, to mitigate its effect and restore it to high fidelity.

This paper is structured as follows: Section \ref{sec:noise} describes the noise model and optimal control equations. Section \ref{sec:results} details the control process of gates; the results are discussed in Section \ref{sec:discussion} and summarized in the concluding Section \ref{sec:conclus}.

\section{Quantum noisy dynamics}
\label{sec:noise}

The noisy dynamics are modeled within open quantum systems \cite{breuer2002theory}.
Specifically, we assume a quantum dynamical semi-group describing Markovian dynamics with a generator in the GKLS master equation form \cite{lindblad1976generators,gorini1976completely}:
\begin{equation}
 \frac{d}{d t} \hat{\boldsymbol{\rho}}(t)=\mathcal{L} \hat{\boldsymbol{\rho}}=-\frac{i}{\hbar}[\hat{\mathbf{H}}, \hat{\boldsymbol{\rho}}]+{\mathcal D} \hat{\boldsymbol{\rho}}\\
   \label{eq: von Neumann equation}
\end{equation}
where $\hat{\mathbf{H}}$ is the system's Hamiltonian, and ${\cal D}$ is the dissipative generator.  $\mathcal{L}$ and $\mathcal{D}$ are termed 
superoperators as they map operators in the Hilbert space to other operators.

\subsection{The noise model}

We consider a quantum system controlled by an external field. The Hamiltonian of the system is:
\begin{equation}
    \hat{\mathbf{H}}  ~=~ \hat{\mathbf{H}}_0+\varepsilon(t) \hat{\mathbf{H}}_c 
\end{equation}
where $\hat{\mathbf{H}}_0$ is the drift Hamiltonian, $\hat{\mathbf{H}}_c$ is the control Hamiltonian, and $\varepsilon(t)$ describes the time-dependent control field.
We aim to study the effect of noise originating from the controller (the pulse generator).
We differentiate between two control noise models: amplitude and phase noise.
For the amplitude noise, we consider random fluctuations
in the control field amplitude { \trr{$\varepsilon(t)=\varepsilon_c(t)(1+\xi(t))$ }}
Assuming the controller is fast relative to the system dynamics, we choose to model the noise as a Gaussian random process
$\langle \xi(t)\xi(t')\rangle= \sqrt{2 \gamma} \delta(t-t')$. Under these conditions, the 
dissipative part of Eq. (\ref{eq: von Neumann equation}) is generated from $\hat{\mathbf{H}}_c$ \cite{gorini1976n,feldmann2010minimal,ruschhaupt2012optimally,levy2018noise,kiely2021exact}:
 \begin{equation}
\begin{array}{l}
{\cal D}_A =  -{\gamma}_A\varepsilon_c^{2}(t)[\hat{\mathbf{H}}_c,[\hat{\mathbf{H}}_c,\bullet]]\\
\end{array}\
\label{eq:ampl}
 \end{equation}
where $\gamma$ is the noise strength.
The phase noise originates from timing errors in the controller \cite{kosloff2010optimal,xuereb2023impact,ball2016role} or frequency fluctuations  \cite{PhysRevLett.37.1383}. Assuming a Gaussian white noise model, the dephasing generator becomes \cite{kosloff2010optimal}:
 \begin{equation}
 \label{eq:phasen}
\begin{array}{l}
{\cal D}_P =  -{\gamma}_P[\hat{\mathbf{H}},[\hat{\mathbf{H}},\bullet]]\\
\end{array}\
 \end{equation}
In both cases, the dissipator depends on the control field.

\subsection{Vectoraization of the Louivile space}
\label{subsec:veccing}

 Obtaining high fidelity of the gates demands the use of a high-fidelity propagation method. We thus use the recently developed semi-global propagation method \cite{schaefer2017semi}, described in Appendix \ref{appendix:prop}. For that purpose, we must define the primary action as a matrix-vector operation in Liouville space. This implies that the operation is a superoperator acting on an operator. Consequently, it is necessary to vectorize Liouville space. All subsequent equations will be described within this framework.

We describe the dynamics employing the Hilbert space of system operators defined by the scalar product.

$\left(\hat{\boldsymbol{A}}\cdot \hat{\boldsymbol{B}} \right) = \text{tr} \{ \hat{\boldsymbol{A}}^{\dagger} \hat{\boldsymbol{B}} \}$.
Superoperators map operators into the same Hilbert space. Hence, we choose to describe operators as vectors  and superoperators 
by matrix-vector multiplications. 
We can now decompose a general operator $\hat{ \boldsymbol{X}}$ employing a full orthogonal operator basis,
\begin{equation}
        \{\hat{\boldsymbol{A}}\}=[{\hat{\boldsymbol{A}}_1},\hat{{\boldsymbol{A}}_2},....,\hat{{\boldsymbol{A}}_N}]
     ~~~~~~~,~~~~
        \text{tr} \{{\hat{\boldsymbol{A}}_i}^{\dagger} \cdot \hat{\boldsymbol{A}}_j \}=\delta_{ij}\nonumber
\end{equation}
Using this basis, we can represent $\hat{\boldsymbol{X}}$ as;
\begin{align}
        \hat{\boldsymbol{X}}=\sum_i \chi_i \hat{\boldsymbol{A}}_i
\end{align}
where $\chi_i$ is an element of a vector $\vec x$ of size $N^2$ expansion coefficients that characterizes $\hat{\boldsymbol{X}}$.
Using this notation, we can define any operator of the dimension $N\times N$ with a vector of size $N^2$. 

In this formalism, the mapping of a superoperator $\Lambda \hat{\boldsymbol{X}} =\hat{\boldsymbol{Y}}$ translates to
a matrix-vector multiplication:
$\tilde{ \boldsymbol{ \mathcal G }} \vec \chi =  \vec \upsilon $
where $\hat{\boldsymbol{Y}}=\sum_i \upsilon_i \hat{\boldsymbol{A}}_i$. \\ \\
The dynamical equation Eq. (\ref{eq: von Neumann equation}) can be reduced to a series of matrix-vector multiplications where $\overrightarrow{\rho}$ will be denoted $ \hat{\boldsymbol{\rho}}$ in the vector space and $\tilde{ \boldsymbol{\mathcal L}}$ will denote the superoperator $\mathcal{L}$ (for additional details, see \cite{am2015three}).  This paper will use the notation " $\tilde{}$ " to assign a superoperator in the Louiville space.
The action of the superoperator on the density operator will read as follows:
\begin{align}       \mathcal{L}\hat{\boldsymbol{\rho}} ~\Rightarrow~\tilde{\boldsymbol{  \mathcal{L}}}\overrightarrow{\rho}
\end{align}

The dephasing superoperators,$\cal {D}_A$ and $\cal{D}_P$, are defined by the following equation:
\trr{
\begin{align}       
\tilde{\boldsymbol{\mathcal{ D}}}_A =  -{\gamma}_A\varepsilon_c^{2}(t)( {\tilde {\boldsymbol{\mathcal H}}}'_c {\tilde {\boldsymbol{\mathcal H}}}'_c \bullet)
\\
\tilde{\boldsymbol{\mathcal D}}_P =  -{\gamma}_A( {\tilde {\boldsymbol{\mathcal H}}}' {\tilde {\boldsymbol{\mathcal H}}}' \bullet)
\nonumber
\end{align}
Where ${\tilde {\boldsymbol{\mathcal H}}}'$ and ${\tilde {\boldsymbol{\mathcal H}}}'_c$  correspond to the superoperators $-\frac{i}{\hbar} [\hat{\mathbf{H}},\bullet]$ and $\frac{i}{\hbar} [\hat{{H}}_c,\bullet]$  in Louiville vector space, respectively.}

More details on vectorization are described in Appendix 
\ref{appndix:vector}.

\subsection{The Optimal Control Theory (OCT) of open systems}
\label{subsec:OCT}

Our general task is to mitigate the effect of noise on a quantum gate.
We employ OCT to provide methods to compute such controls. The external fields $\left\{\mathcal{E}_k\right\}$ interacting with the quantum system have the task of steering the system's dynamics from an initial to a final state. 
A more ambitious task 
is to find the driving field that generates a quantum map $\Lambda(T)$ \cite{palao2003optimal}.
The dynamical equation of motion for the map becomes:
\begin{equation}
   \frac{d\Lambda(t)}{dt}= \mathcal{L}(t) {\Lambda}(t) ~~\Rightarrow ~~
   \frac{d \tilde{ \boldsymbol{ \mathcal G}}}{dt}=\tilde{ \boldsymbol{\mathcal  L}}(t) 
   \tilde{ \boldsymbol{ \mathcal G}}, 
   \label{Vecor-Louivile}
\end{equation}
where $\tilde{ \boldsymbol{ \mathcal G}}$ is the evolution operator in the vector space. 
The initial condition is $\Lambda(t=0)=\mathcal{I}$, where $\mathcal{I}$ is the identity superoperator. 
The objective is to obtain the optimal driving field $\epsilon(t)$ that induces a given transformation ${\mathcal O}$ at $t=T$ 
($ {\mathcal O}\Rightarrow \tilde{ \boldsymbol{\mathcal  O}}$).
This can be interpreted as the 
mapping of a complete set of operators $\{ \hat{\boldsymbol{A}} \}$ as close as possible to the desired  map ${\mathcal O}$.

In OCT, this task is translated to optimizing the objective functional \cite{petruhanov2023quantum}:
\trr{
\begin{eqnarray} 
\begin{array}{c}
 \mathcal{J}_{max}= \text{Tr}\{{\mathcal{O}}^{\dagger} \Lambda (T) \} 
 = \sum_j \text{tr} \{({\mathcal{O}} 
 \hat{\boldsymbol{A}}_j )^{\dagger}
 ( \Lambda (T) \hat{\boldsymbol{A}}_j) \} 
 \Rightarrow \text{Tr} \{ \tilde {\boldsymbol{{\mathcal{O}}}} \tilde{ \boldsymbol{ \mathcal G}}  \}
 \end{array}
\end{eqnarray}}
 
Where $\tilde{\boldsymbol{ \mathcal G}}$ corresponds to $\Lambda$ in the Hilbert space. We use the symbol $\text{Tr}$ to denote a trace of superoperators, while $\text{tr}$ 
denotes the trace operators in the Hilbert space.
Two constraints are added to the objective. The first restricts the dynamics to comply with the Liouville-von Neumann equation Eq. (\ref{Vecor-Louivile}), 
\begin{equation}
   {\mathcal{J}_{con}}= \int_0^T \operatorname{Tr}\left\{\left(\frac{\partial \Lambda(t)}{\partial t}- {\mathcal{L}}(t) \Lambda(t)\right) \Upsilon(t)\right\} dt
   \label{eq:constrainfunct}
\end{equation}
where $\Upsilon(t)$ is a superoperator Lagrange multiplier.
The second constraint restricts the total field energy
\begin{equation}
   \mathcal{J}_{penal} =\lambda \int_0^T \frac{1}{s(t)}|\epsilon(t)|^2 d t
   \label{eq:limpower}
\end{equation}
$\lambda$ is a scalar Lagrange multiplier, and $s(t)$ is a shape function that smoothly turns the pulse on and off. For this task, we choose a Gaussian profile.
Adding these constraints, we obtain the overall functional,
\begin{equation}
   \mathcal{J}_{Tot}= \mathcal{J}_{max}+\mathcal{J}_{con}+\mathcal{J}_{penal}
   \label{eq:functional}
\end{equation}

The control task is translated to the maximization of the generalized objective ${\mathcal{J}_{Tot}}$, $\delta {\mathcal{J}_{Tot}}=0$.
Functional derivatives with respect to the various functional elements are then taken, $\Upsilon, \Lambda$, and $\epsilon$, resulting in the following set of control equations:
\begin{enumerate}
\item 
The Louivile equation Eq. (\ref{Vecor-Louivile}) with the initial condition $\Lambda(t=0)=\mathcal{I}$ for $\Lambda$.
\item
The inverse Louivile equation
\begin{equation}
   \frac{\partial \Upsilon(t)}{\partial t}= \mathcal L^* (t) \Upsilon(t) ~~\Rightarrow ~~
   \frac{\partial \tilde{ \boldsymbol Y}(t)}{\partial t}=\tilde{ \boldsymbol{\mathcal L}}^{\dagger}(t) \tilde Y(t)
   \label{eq:liouvil-control}
   \end{equation}
where $\tilde{ \boldsymbol Y}$ corresponds to $\Upsilon$ in the vector space, \trr{ a functional derivative with respect to time results in the initial condition for the reverse propagation of the Lagrange map multiplier $$\Upsilon(t=T)={\mathcal{O}}^{\dagger}$$ \cite{palao2003optimal}}. Note that the loss of information in the reverse equation is also inverted so that dephasing occurs from both time ends. 
\item 
The control field updated the equation for both amplitude and phase noise cases:
\begin{equation}
    \begin{aligned}
          \Delta \boldsymbol{\epsilon}(t)=-\operatorname{Im}\frac{s(t)[\operatorname{Tr} \{ \Upsilon(t) {\boldsymbol{\mathcal H}}'_c \Lambda(t)\}]}{2 (\lambda+{\gamma}_A[\operatorname{Tr} \{ \Upsilon(t) {{\boldsymbol{\mathcal H}}'_c}^2 \Lambda(t)\}])}  ~~\Rightarrow~~ -{s(t)}\operatorname{Im}\frac{\text Tr\{\tilde{\boldsymbol Y}^* \tilde {\boldsymbol{\mathcal H}}'_c \tilde{\boldsymbol{ \mathcal G}}(t)\}}{2( \lambda+ {\gamma}_A Tr\{\tilde{\boldsymbol Y}^* \tilde{\boldsymbol{ \mathcal H}}_c^2 \tilde{\boldsymbol{ \mathcal G}}(t)\})}  
    \end{aligned}
    \label{eq:update}
\end{equation}
where:
$\frac{\partial}{\partial \epsilon} \tilde {\boldsymbol{\mathcal L}}
= {\tilde {\boldsymbol{\mathcal H}}}'_c$ and 
$\frac{\partial}{\partial \epsilon} \tilde {\boldsymbol{\mathcal D}_A}
=2 {\gamma}_A\varepsilon_c \tilde{\boldsymbol{ \mathcal H}}_c^2 $
\end{enumerate}
The dynamical equations are solved iteratively (counted by k), employing the Krotov method \cite{krotov1995global,palao2003optimal}, leading to an update to $\boldsymbol{\epsilon}^{(k)}$:
\begin{equation}
    \begin{aligned}
       \Delta \boldsymbol{\epsilon}^{(k)}(t)=-\frac{s(t)}{2 \lambda} \operatorname{Im}\left[\operatorname{Tr}\left\{\Upsilon^{(k-1)}(t) \mathcal L_c\Lambda^{(k)}(t)\right\}\right]  \Rightarrow -\frac{s(t)}{2 \lambda} \operatorname{Im}\left[\text Tr\left\{\tilde{\boldsymbol  Y}^{(k-1)}(t)  {\boldsymbol{\mathcal H}}'_c\tilde{\boldsymbol{ \mathcal G}}^{(k)}(t)\right\}\right].
    \label{eq:update}
    \end{aligned}
\end{equation}
This evaluation continues until absolute fidelity $(1-\mathcal{J}_{max})$ is reached. Note that for most cases $\lambda \gg \gamma_A$, one can neglect its term to obtain the denominator in Eq. (\ref{eq:update}). However, for high fidelity values, especially close to the stagnation limit of the Krotov iterations, it was found that the term could improve the outcomes significantly.

For the high-fidelity gates employed here, one needs a high-fidelity propagator. We thus solve numerically the dynamics in Eq. (\ref{Vecor-Louivile}) and (\ref{eq:liouvil-control}), using the semi-global propagation method \cite{schaefer2017semi}, described in Appendix \ref{appendix:prop}.
With these solutions, we use Eq. (\ref{eq:update}) to update the field and repeat until
the desired fidelity is reached. The Krotov method is known to stagger after many iterations. Therefore, we halt this iterative process when stagnation is reached.

\section{\trr{Control of quantum gates}}
\label{sec:results}

OCT is employed first
to obtain the desired unitary gate
without dissipation.
This solution serves as a reference
for studying the effect of noise on fidelity. At this point, optimal control 
is used again, including the dissipation to search for control fields that mitigate
the impact of noise.
The generator of the dynamical map has the form:
\begin{equation}
    \tilde{\boldsymbol{ \mathcal L}}={\boldsymbol{\mathcal H}}'_0+ {\boldsymbol{\mathcal H}}'_c+ \tilde{ \boldsymbol{\mathcal D}}_{A/P}.
    \label{total dynamical map}
\end{equation}
where $\tilde{ \boldsymbol{\mathcal D}}$ generates noise, and the indices '$A$' and '$P$' represent the two types of amplitude and phase quantum noise defined in Eq. (\ref{eq:ampl}) and Eq. (\ref{eq:phasen}), respectively.

We choose a driven system that is completely unitary and controllable. This controllability results from commutators of the drift Hamiltonian, ${\boldsymbol{\mathcal H}}'_0$, and the control Hamiltonian, ${\boldsymbol{\mathcal H}}'_c$, which generate the full-rank algebra \cite{ramakrishna1996relation}. 
Three representative target gates are studied:
\begin{enumerate}
\item{
     A Hadamard gate on a single-qubit SU(2)}
     
\item{Pauli-X gate - SU(2) embedded in SU(4).}

\item{
    An entangling gate on a 2-qubit system SU(4).}
\end{enumerate}

The generator of the dynamical map is Eq. (\ref{total dynamical map}), and the cases differ in their Hamiltonians. Consequently, the dephasing elements of the different instances represented by Eq.'s (\ref{eq:ampl})  and (\ref{eq:phasen}) differ. 

The first step in OCT is to obtain the reference unitary gate,  which also establishes a good guess of the control field $\varepsilon(t)$. A good guess 
can become essential for convergence to high fidelity. \trr{In many cases, the target is "hard to find," and a good guess initiates a fast convergence rate to the solution\cite{kallush2011scaling}.} In this case, the OC  dynamics is generated by:
\begin{equation}
 \tilde{\boldsymbol{ \mathcal L}}=\tilde{ \boldsymbol{\mathcal H}}_0+ \tilde{ \boldsymbol{\mathcal H}}_c.
    \label{No-noise-L}
\end{equation}
We define fidelity as $F = \text{Tr}\{\hat{\mathrm{O}}^{\dagger} \Lambda (T)\}$ and infidelity as $IF = 1-\text{Tr}\{\hat{\mathrm{O}}^{\dagger} \Lambda (T) \} $. The unitary infidelity is assigned by $IF_U$.

To test the effect of noise on the infidelity, we use the solution obtained from $IF_U$. We propagate a system with noise, using Eq. (\ref{total dynamical map}) as the generator of the dynamical map.

\begin{figure*}
\vspace{-3cm}

\centering
    \includegraphics[scale=0.47,angle =-90]{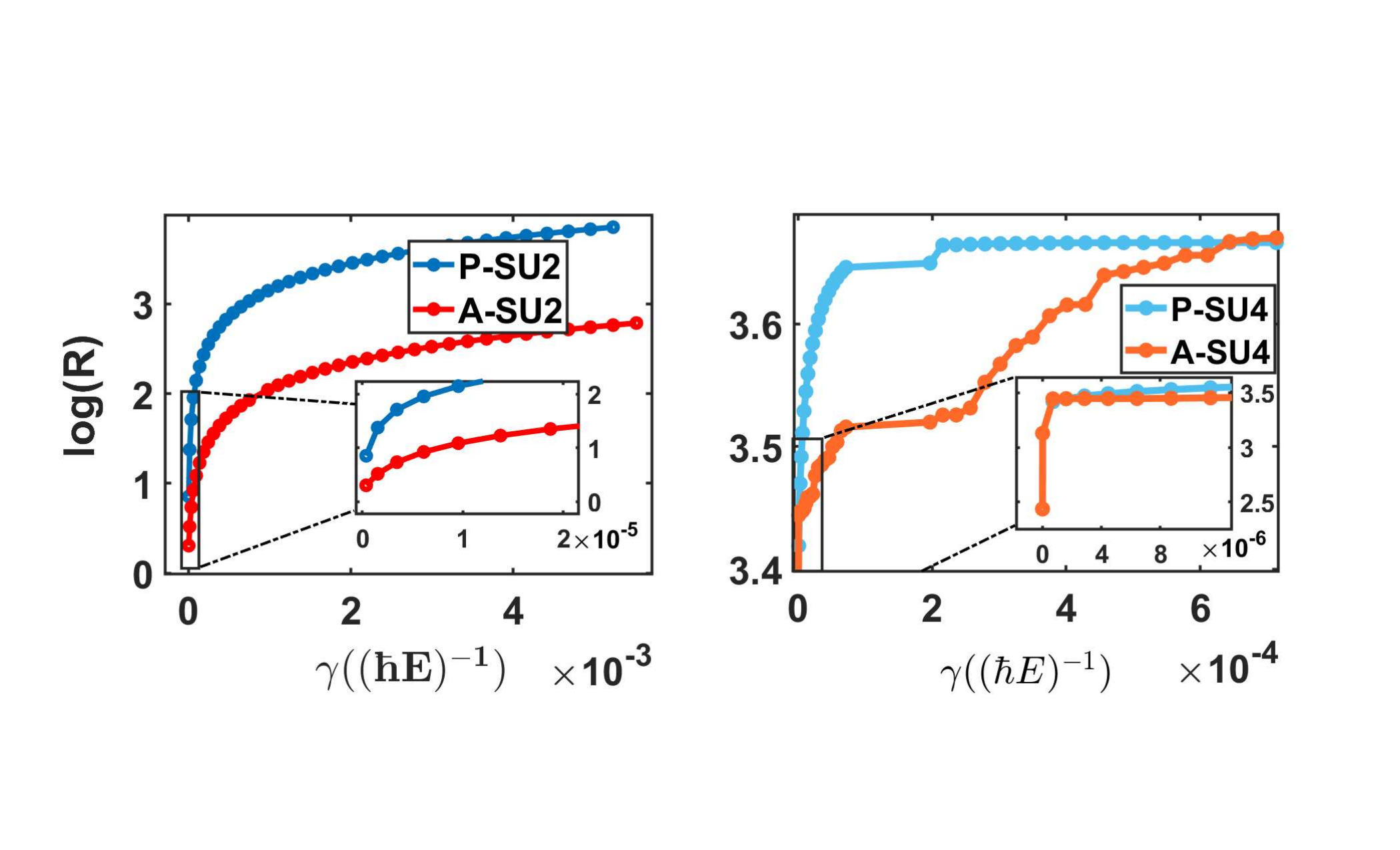}
    \vspace*{-15mm}
    \caption{\textbf{Destruction of fidelity by noise.}
    The change in the infidelity of a quantum gate $IF = 1-\text{Tr}\{\hat{\mathrm{O}}^{\dagger} \Lambda (T) \} $. The reference is a converged unitary control without noise with infidelity $IF_U$. The same field is used to obtain the gate with noise $IF_n$.
    The log of the ratio $R=\frac{IF_n}{IF_U}$,
    $\log (R)= \log (IF_n)- \log (IF_U)$, is displayed as a function of the noise strength $\gamma$. Both phase and amplitude noise are displayed. The insert displays a zoom on the low noise level.
    Left panel: The single qubit Hadamared gate.
    Right panel: The two-qubit entangling gate.} 
    \label{fig:noise-distruction}
\end{figure*}

The degradation of the fidelity by the noise from the infidelity of $IF_U$ to $IF_n$ is presented in Fig. \ref{fig:noise-distruction}. We plot the log of the ratio $R=\frac{IF_n}{IF_U}$ versus the noise rate $\gamma$.

Fig. \ref{fig:noise-distruction} shows that the system is more resilient to amplitude noise $\tilde D_A$ than to phase noise $\tilde D_P$ for the Hadamard gate. The degradation is significant when the noise strength increases, saturating with three orders of magnitude in infidelity.  
The two-qubit entangling gate is two orders of magnitude more sensitive than the single-qubit gate with respect to the noise strength $\gamma$.

\subsection{Hadamard gate-SU(2) space}
\label{subsec:hadamard}

The Hadamard gate performs a rotation of $\pi$ about the axis $(\hat{x}+\hat{z}) / \sqrt{2}$ of the Bloch sphere. 
Expressed as a superoperator in  the operator basis $\hat I, \hat S_X,\hat S_Y,\hat S_Z$, the Hadamard gate is given by:
\begin{equation}
\Lambda_U = 
\left(
\begin{array}{cccc}
1 & 0 & 0 & 0\\
0 & 0 & 0 & -1 \\
0 & 0 & -1 & 0 \\
0 & -1 & 0 & 0 
\end{array}
\right)~~.
\label{hadtran}
\end{equation}

\begin{figure}
     \vspace{-0.9cm}
    \centering    
    \includegraphics[scale=0.50,angle =-90]{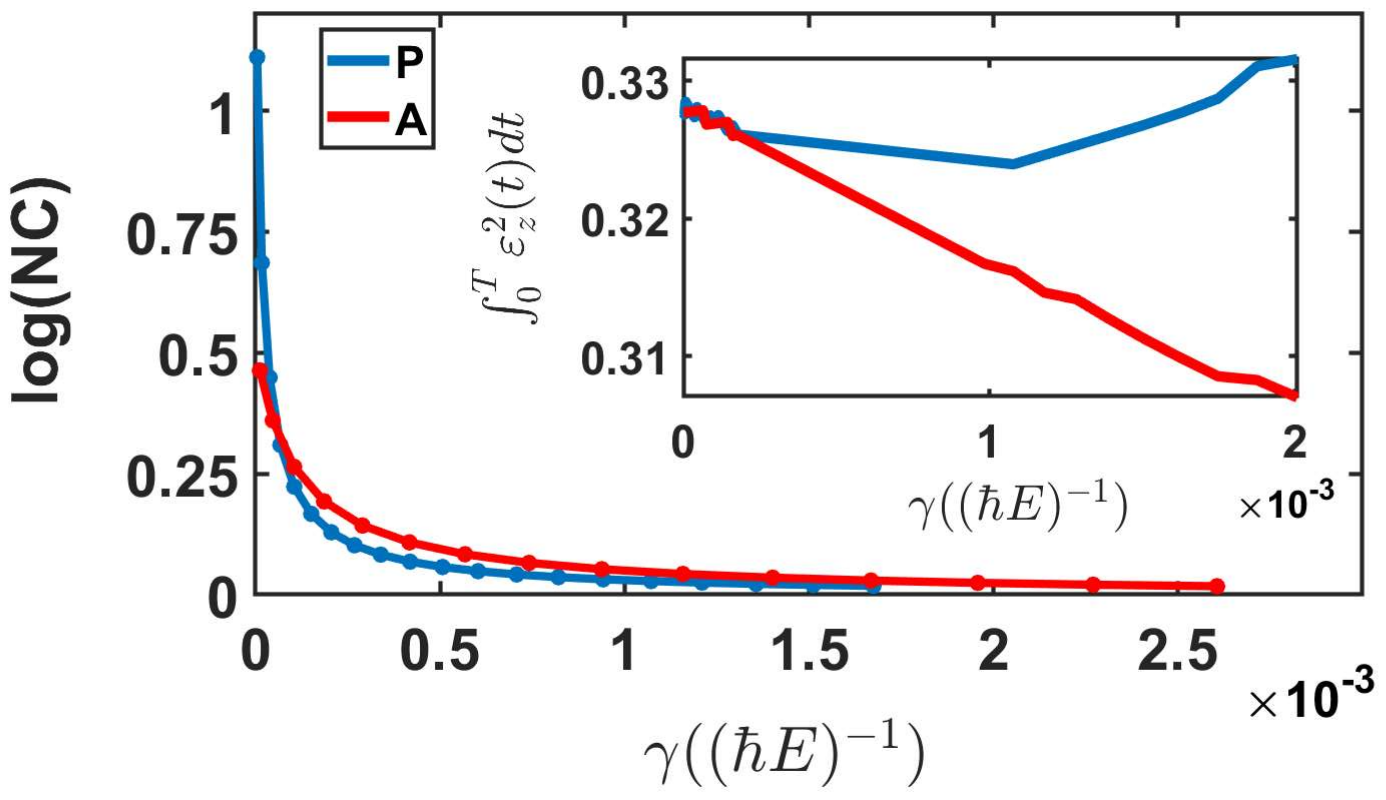}
     \vspace*{-20mm}
    \caption{\textbf{Reduction of the noise impact via Optimal Control for the Hadamard gate at short times (2-Rabi cycles):} The reduction impact is defined as $NC=\frac{IF_n}{IF_F}$ vs. the dephasing rate, $\gamma$ in Logarithmic scale. 
    The two dephasing mechanisms, amplitude noise, $\tilde D_A$, (red, {A}), and phase noise, $\tilde D_P$, (blue, {P}), are presented.
    (inset) The accumulated energy of the control field vs. the dephasing rate in the two dephasing mechanisms.}
    \label{fig:Delta-G-2-Had}
     \vspace{-8mm}
\end{figure}

We chose a Hamiltonian, which allows high fidelity for this type of target gate. The drift Hamiltonian is:
\begin{equation}
   \hat{\mathbf{H}}_0= u{\hat S}_Z+a_X{\hat S}_X
    \label{free-dynamics-hadamard-1}
\end{equation}
And the control Hamiltonian:
\begin{equation}
    \hat{\mathbf{H}}_c = \varepsilon(t){\hat S}_{XY}
        \label{control-dynamics-hadamard-1}
\end{equation}
where ${\hat S}_{XY}$ is a rotation w.r.t $XY$ plane (any direction in this plane leads to control).
Using this generator, we first obtain the reference unitary gate by OCT, with unitary infidelity $IF_U = 1\times 10^{-5}$.
Using the optimal unitary control field as a guess,
solutions under noise are obtained as a function of $\gamma$, noise rate.
For each value of $\gamma$, we apply OCT
to mitigate the effect of noise.
Fig. \ref{fig:Delta-G-2-Had} depicts the infidelity improvement as a function of the noise rate, $\gamma$, for a short control period (2-Rabi cycles $\tau=4 \pi/\Omega $ ~,~$ \Omega=\sqrt{u^2 + a_X^2} $). 

The improvement ratio is defined as $NC=\frac{IF_n}{IF_F}$, where $IF_n$ is the infidelity obtained using the unitary guess field subject to noise, and $IF_F$ is the infidelity brought by optimal control mitigating the noise. \trr{Under this definition, $IF_F$ starts with the fidelity of the initial guess achieved for $IF_n$ and can only be improved due to Krotov's method of monotonic convergence. Its minimal value is hence $IF_n$, so the minimal value of $log(NC)$ is zero in the worst case.} 

Infidelity can be mitigated with OC by a factor of two at low noise rates ($\gamma<$0.2$\times 10^{-3}$) for phase noise and on amplitude noise systems in high noise rate ($\gamma>$0.2$\times 10^{-3}$) systems. The optimal control formulation restricts the accumulated power. The field intensity is optimized without noise. Nevertheless, we observe that the invested pulse energy in the optimal solution varies slightly with the noise strength $\gamma$. The inset of Fig. \ref{fig:Delta-G-2-Had} shows this dependence. Especially for amplitude noise, where the source of decoherence is the field strength, optimal control reduces the field strength and, consequently, the controller's noise to tackle phase noise at slow noise rates. 
However, more substantial noise causing significant dephasing can only be controlled by boosting the field's strength. Thus, for phase noise (${\mathcal D}_P$), the OCT solution turns on the field's strength at elevated noise levels 
(Cf. Eq. (\ref{eq:purityloss})~). \trr{ In appendix \ref{Appendix E}, we show examples of typical control fields, with or without noise. Small changes in the field have a significant effect on fidelity \cite{aroch2023employing}.}

\begin{figure}[h!]
\hspace*{-1.2cm} 
    \includegraphics[scale=0.42,angle =-90]{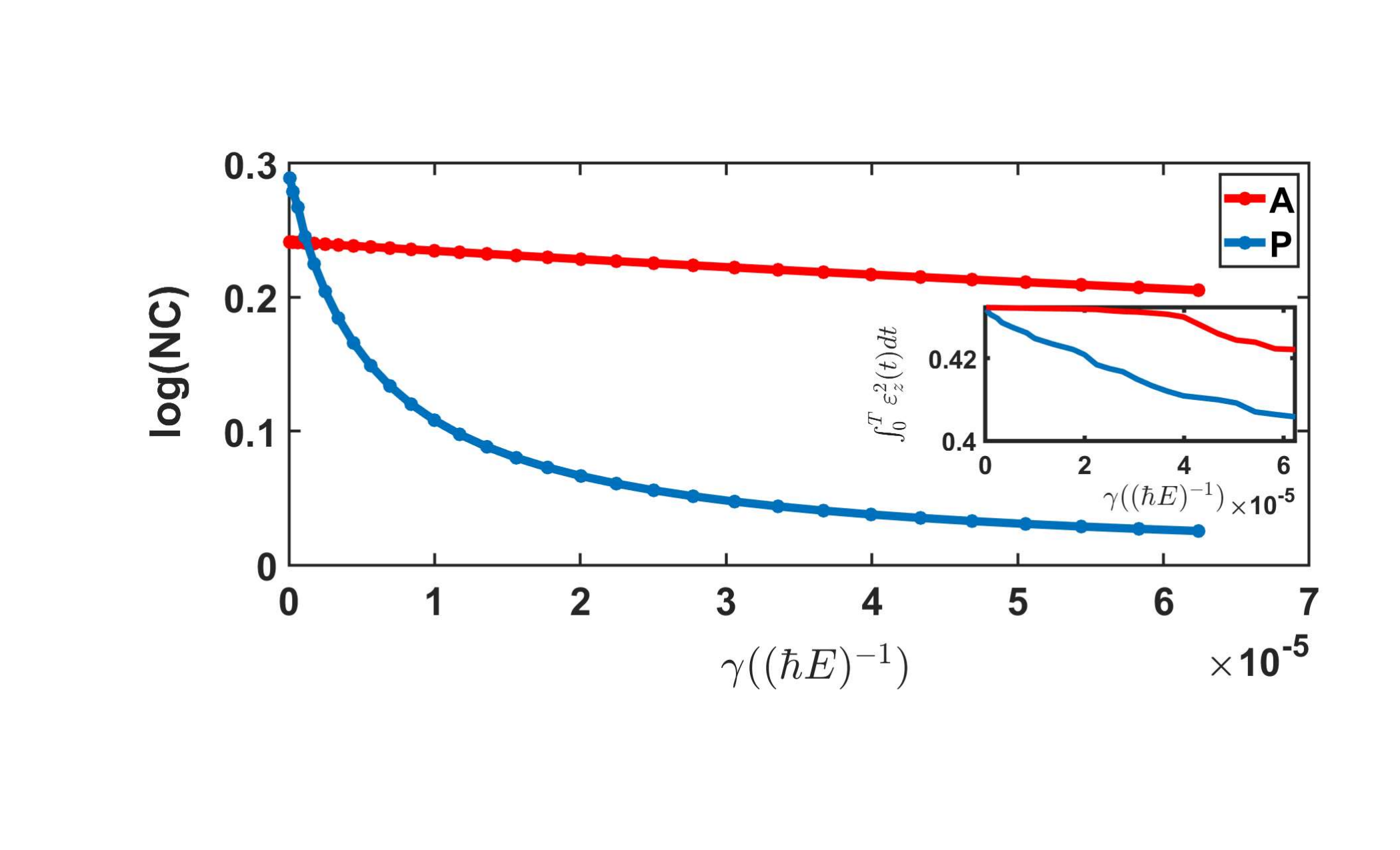}
    \vspace{-1.9cm}
\caption{\textbf{Noise reduction using Optimal Control for the Hadamard gate at longer times (6-Rabi cycles):}
The two dephasing mechanisms, 
amplitude noise, $\tilde D_A$, (red, {A}) and phase noise, $\tilde D_P$ (blue), (blue, {P}), are presented.
(inset) The accumulated energy of the control field vs. the noise rate in the two noise mechanisms.}
    \label{fig:Delta-G-6-Had}
\end{figure}\par

Fig. \ref{fig:Delta-G-6-Had}
shows the infidelity improvement 
for a longer control period (6-Rabi-cycles). When increasing the time allocated for control, the reference infidelity is maintained at $IF_U \sim 1 \times 10^{-5}$. However, to maintain comparable infidelity for both cases, the dephasing rate $\gamma$ must be reduced by two orders of magnitude.
In both cases, we find that OCT mitigates the effect of noise
at low values of $\gamma$. When the dissipation increases, the ability to mitigate the noise vanishes. 

Optimal control mitigation can reduce $IF_F$ for the phase noise more effectively than $IF_F$ of the amplitude noise at low noise rates. This trend is reversed at high noise rates. The inset of Fig. \ref{fig:Delta-G-6-Had} shows that
the optimal control solution reduces the field's energy for higher dephasing rates, resulting in higher fidelity.
\begin{figure}[h!]
\vspace{1.cm}
\centering
\includegraphics[scale=0.44,angle =-90]{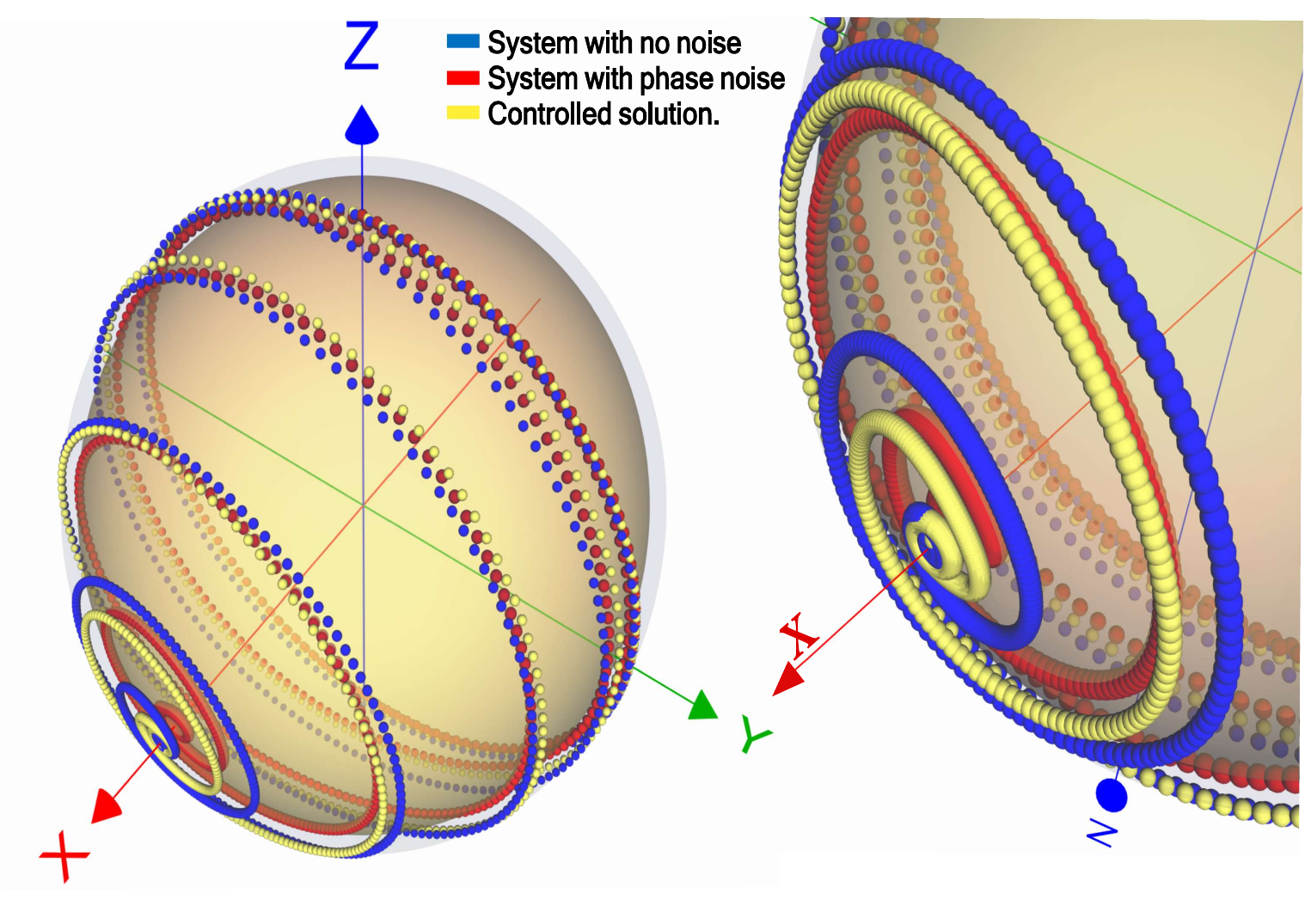}
\caption{\textbf{The trajectory of an initial operator in the $Z$ direction during a Hadamard transformation under phase noise.}
    The reference unitary evolution in blue has an infidelity of $IF_U=1 \times 10^{-5}$. The trajectory with noise, in red, 
    does not reside on the surface of the Bloch sphere. The optimal control solution, in yellow, mitigates the infidelity to a value of $IF_n=15 \times 10^{-3}$.
    The right panel zooms on the target state. The outer sphere represents the purity of 1. The inner-sphere purity is $90\%$.}
    \label{fig:bloch}
\end{figure}

Fig. \ref{fig:bloch} follows a trajectory of an element of the operator basis during the Hadamard transformation. When the generator is noiseless, the dynamics stay on the surface of the Bloch sphere, and noise pushes the dynamics into the interior of the Bloch sphere. Optimal control mitigation helps to redirect the trajectory towards the surface. 

\subsection{Pauli-X gate - SU(2) embedded in SU(4).}

The two-qubit entangling and single-qubit gates comprise a universal computational set that can simulate any quantum circuit. 
We first optimize the control field by generating the single-qubit $Pauli-X$ gate embedded in a two-qubit Hilbert space:
\begin{align}
\hat W=
\left(\begin{array}{llll}
0 & 0 & 0 & 0 \\
0 & 0 & 0 & 0 \\
0 & 0 & 0 & 1 \\
0 & 0 & 1 & 0
\label{Pauli-X}
\end{array}\right)
\end{align}
This gate rotates the second qubit 
on the $X$ axis conditioned on the first qubit upper projection in the $Z$ direction. 
For the lower projection, no restriction is imposed.

For this task, we employ the trivial drift generator:
\begin{equation}
    \tilde{\boldsymbol{ \mathcal H}}_0= \tilde {\cal I}^1\otimes \tilde {\cal I}^2
    \label{free-dynamics-hadamard}
\end{equation}
And the following control generator:
\begin{equation}
    \tilde{\boldsymbol{ \mathcal H}}_c= \varepsilon(t)\sum_{i=X,Y}a_i(\tilde {\cal I}^1-{\tilde S^1}_Z) \otimes {\tilde S^2}_i~~,
        \label{control-dynamics-hadamard-2}
\end{equation}
It is sufficient to generate the unitary gate. This framework is then employed to study
the two noise models, ${\mathcal D}_A$ and ${\mathcal D}_P$. 

First, we employ OCT on pure unitary dynamics to obtain a high-fidelity solution converging to infidelities of $IF_U ~\sim 1 \times 10^{-5}$.

Next, we employ this solution as a guess for the OCT protocol on noisy systems, mitigating noise degradation.
\begin{figure}
\includegraphics[scale=0.45]{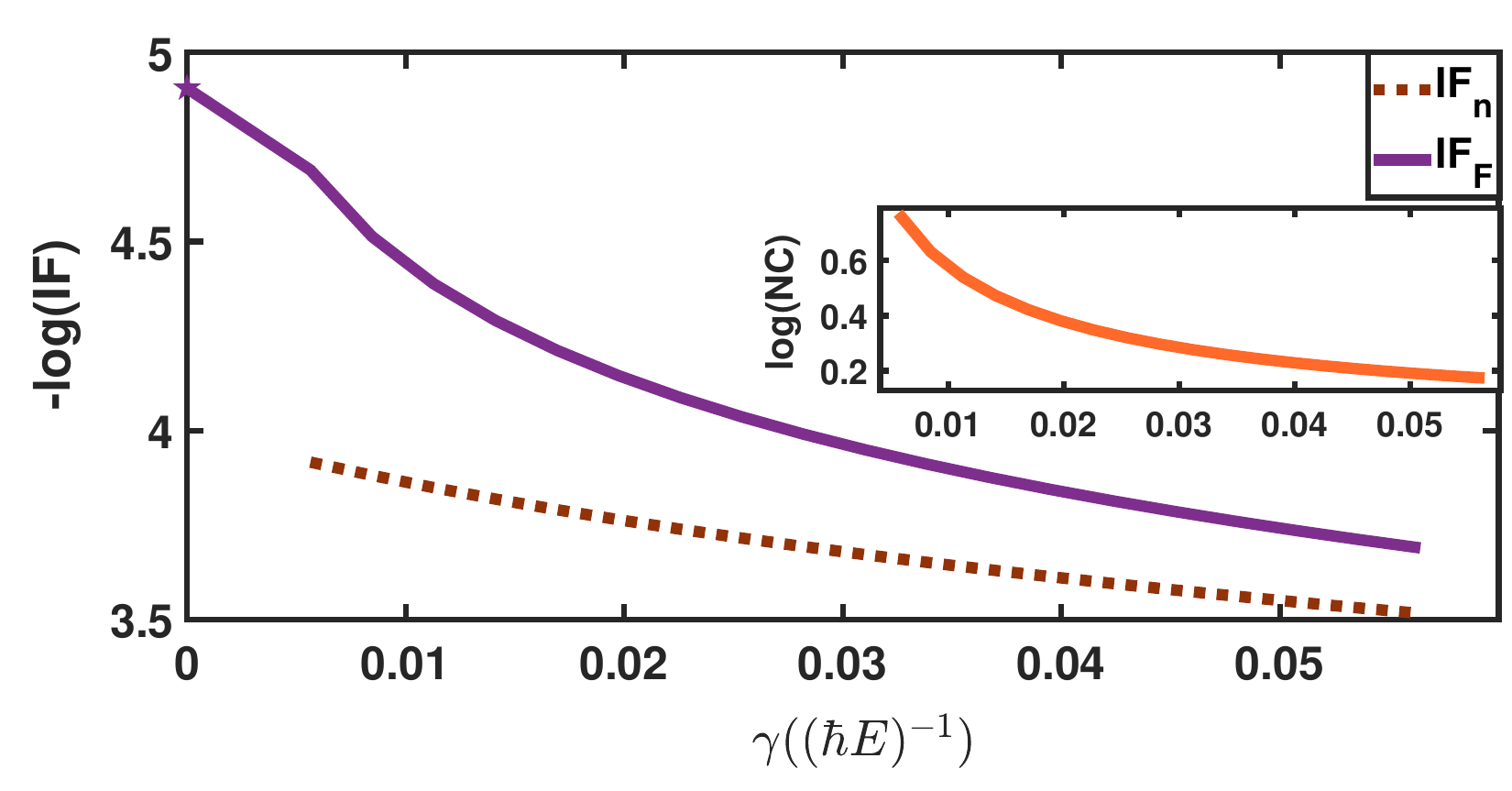}
    \vspace*{0mm}\caption{\textbf{
    Infidelity as a function of the noise strength $\gamma$ for the Pauli-X gate.} 
    The dashed line (brown, $IF_n$):  infidelity
employing the unitary control field. The solid line (purple, $IF_F$): infidelity 
of the OCT solution, mitigating noise.
(inset) The noise reduction, NC (log scale), vs. ,$\gamma$. }
    \label{fig:Delta-G-SU2-CNOT}
\end{figure}

Fig. \ref{fig:Delta-G-SU2-CNOT} shows the degradation due to noise $IF_{n}$ 
relative to $IF_U$ (star)
and the mitigation due to OCT ($IF_{F}$).
The inset quantifies the noise mitigation due to OCT. 
In the previous example, we examined the control of a single qubit embedded in a two-qubit space but uncoupled from it. 
\\
Here, we add the coupling between the two qubits. \trr{To reach an adequate initial guess for the entangled two-qubit gate, we employ the following strategy: First, the single qubit objective is extended to control a single qubit in a noisy environment, with the second qubit acting as an ancilla. Then, the optimized field is employed as an initial guess for the fully entangled two-qubit gate.}
In the first stage, the objective remains the same as Eq. (\ref{Pauli-X}).
Then, we modify the drift Hamiltonian, removing the resonance condition between the two qubits: 
\begin{equation}
    \tilde{\boldsymbol{\mathcal  H}}_0= 
    a \tilde {\cal{I}}^1\otimes \tilde{\cal{I}}^2 +\omega_1 
    {\tilde S}_Z^1\otimes \tilde{\cal{I}}^2
    \label{free-dynamics-hadamard3}
\end{equation}
where $a$ is a phase factor, and $\omega_1$ is the qubit frequency. Further, we increase the number of control fields to include additional coupling:
\begin{equation}
    {{^Z}\tilde{\boldsymbol{\mathcal H}}_c}= \varepsilon_Z(t)\sum_{i=X,Y}a_i(\tilde {\cal I}-{\tilde S}_Z)^1 \otimes {\tilde S}_i^2 
        \label{control-dynamics-CNOt_E_1}   
\end{equation}
\begin{equation}
    {^E}\tilde{\boldsymbol{ \mathcal H}}_c= \varepsilon_E(t)({\tilde S}_X^1 \otimes {\tilde S}_Z^2)
        \label{control-dynamics-CNOt_E_2}
\end{equation}
Here, the field ${^E}\tilde{\boldsymbol{ \mathcal H}}_c$ adds a correlation between the qubits.

\begin{figure}
\vspace{-.0cm}
	\centering
\includegraphics[scale=0.61]{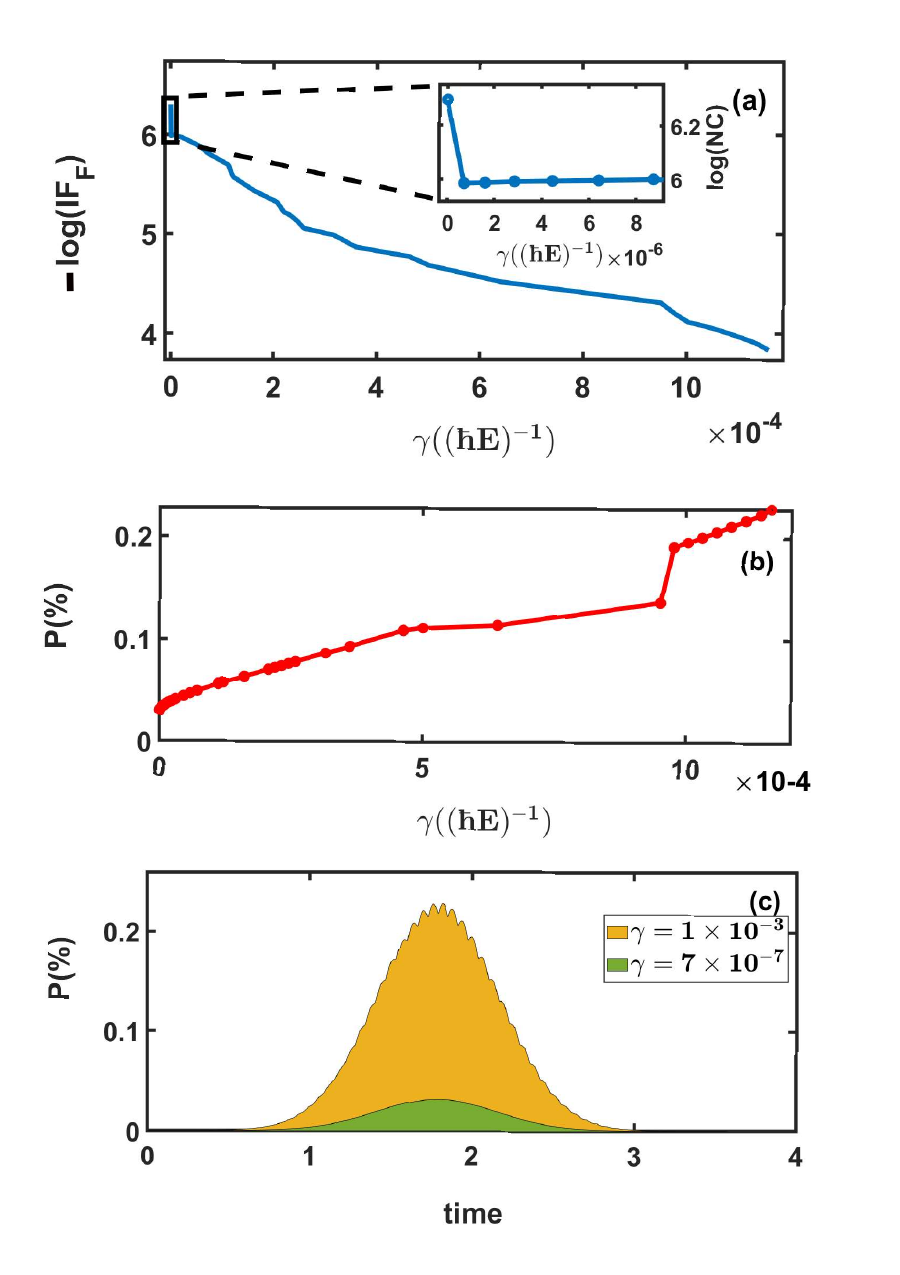}
\vspace{-.5cm}
\caption{\textbf{Single qubit gate with an ancilla: a) Noise reduction due to optimal control as a function of dephasing noise.} The insert shows the small dephasing regime. b) The maximum population in the ancilla as a function of the dephasing rate. c) The population in the ancilla as a function of dimensionless time for two noise strengths.
The time is relative to the energy scale of $\hat{\mathbf{H}}_0$: $\Delta t =2 \pi/\Delta E$. }  
    \label{fig:ancila-fig}
\end{figure}

Fig. \ref{fig:ancila-fig}(a) demonstrates the infidelity achieved by OCT as a function of the noise strength.
The inset illustrates the ability of OCT to harness noise to achieve higher fidelity in the mild noise regime. 

To understand how OCT works and reduces noise, we analyze the temporal population in the ancilla.
Fig. \ref{fig:ancila-fig}(b) displays the maximum population (\%) in the ancilla as a function of noise strength. When noise levels increase, we observe that additional population is transferred to the ancilla state. To reach the objective, OCT must return the ancilla population to the qubit. 
Fig. \ref{fig:ancila-fig}(c) shows the time dependence of the transformed population under two distinct noise regimes: high (yellow) and low (green).
Initially, the ancilla states are empty, meaning only the qubit states are occupied. We compare this control problem with the one shown in Fig. \ref{fig:Delta-G-SU2-CNOT}, where the coupling Hamiltonian is different, but the target and the initial state are the same. 

Notice that there are now two sources of dephasing ($\tilde D_P$) noise, one of which couples the system out of the qubit states. 
When introducing higher noise levels, we notice that the coupling to the ancilla affects our system infidelities, making it harder for OCT to achieve its aim. 
However, at low dephasing rates, we can observe that the ancilla helps to achieve lower infidelities. Now, the fields obtained with the ancilla can be used for the two-qubit gate.

\subsection{Entangling gate-SU(4) space}
\label{subsec:su4}

The ultimate test of optimal control mitigation employs the full palette of the 16 operator bases of SU(4). The dynamic generators connect all elements of the algebra using the one described in Eq's. (\ref{free-dynamics-hadamard3}),(\ref{control-dynamics-CNOt_E_1}),(\ref{control-dynamics-CNOt_E_2}).

\begin{figure} 
\includegraphics[scale=0.35,angle =-90]{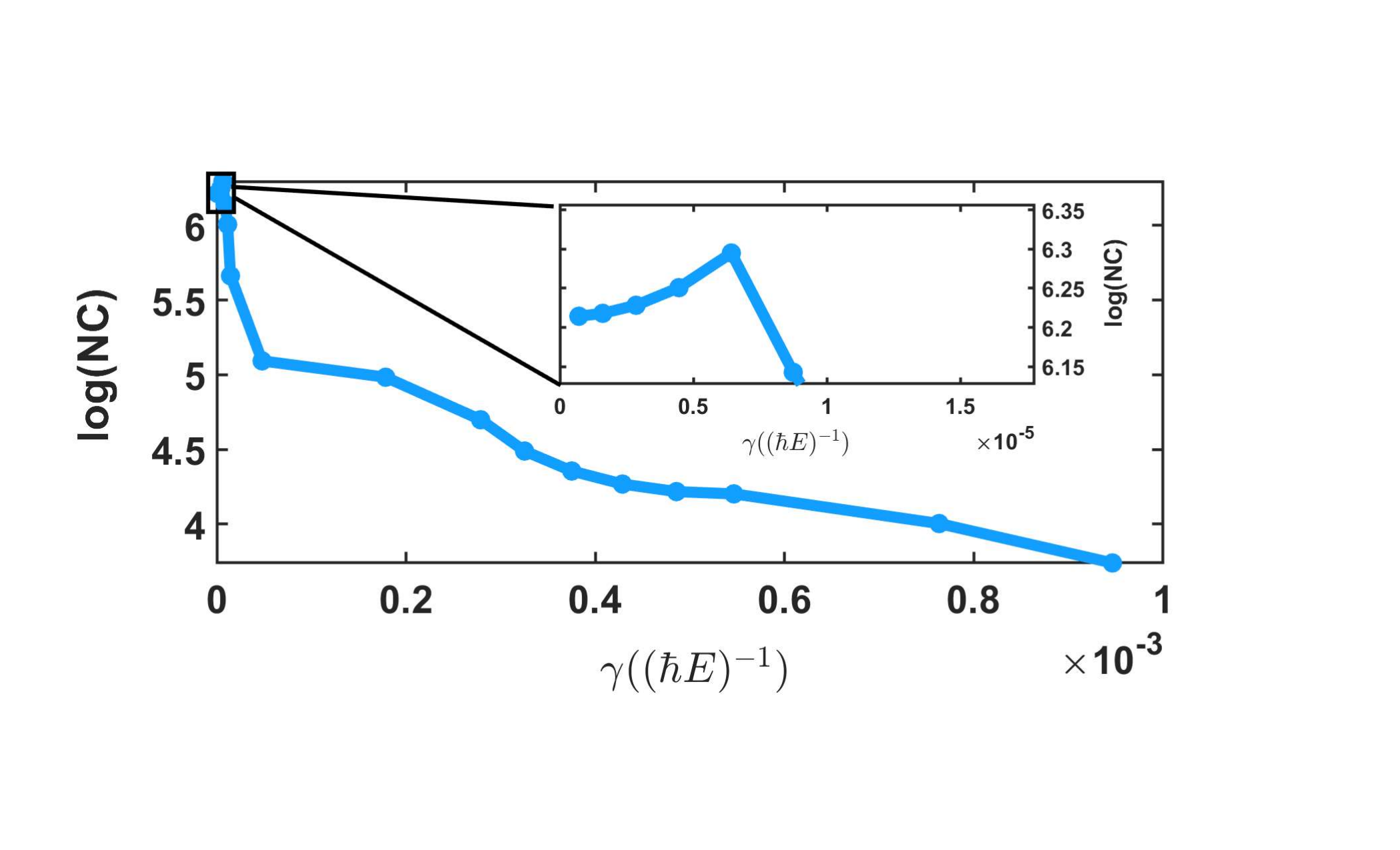}
\vspace{-1.3cm}
\caption{\textbf{The noise cancellation $NC=\frac{IF_n}{IF_F}$ (log scale) Vs. the noise rate $\gamma$ for the entangling gate under phase noise. The insert focuses on the low noise regime.}} 
\label{fig:Delta-G-H-SU4-CNOT}
\end{figure}

Obtaining converged results is challenging. The entangling gate set as a target is:
\begin{align}
\hat U=
\left(\begin{array}{llll}
1 & 0 & 0 & 0 \\
0 & 1 & 0 & 0 \\
0 & 0 & 0 & i \\
0 & 0 & -i & 0
\end{array}\right)
\end{align}
The CNOT+Phase gate can be 
part of the universal set.
We, therefore, employ it to test the effect of noise.

OC can reduce the infidelity to $IF_U \sim 1 \times 10^{-6}$ for the unitary case when the Krotov iterative process reaches stagnation.

\begin{figure}
\vspace{-2.46cm}
\centering    \includegraphics[scale=0.35,angle =-90]{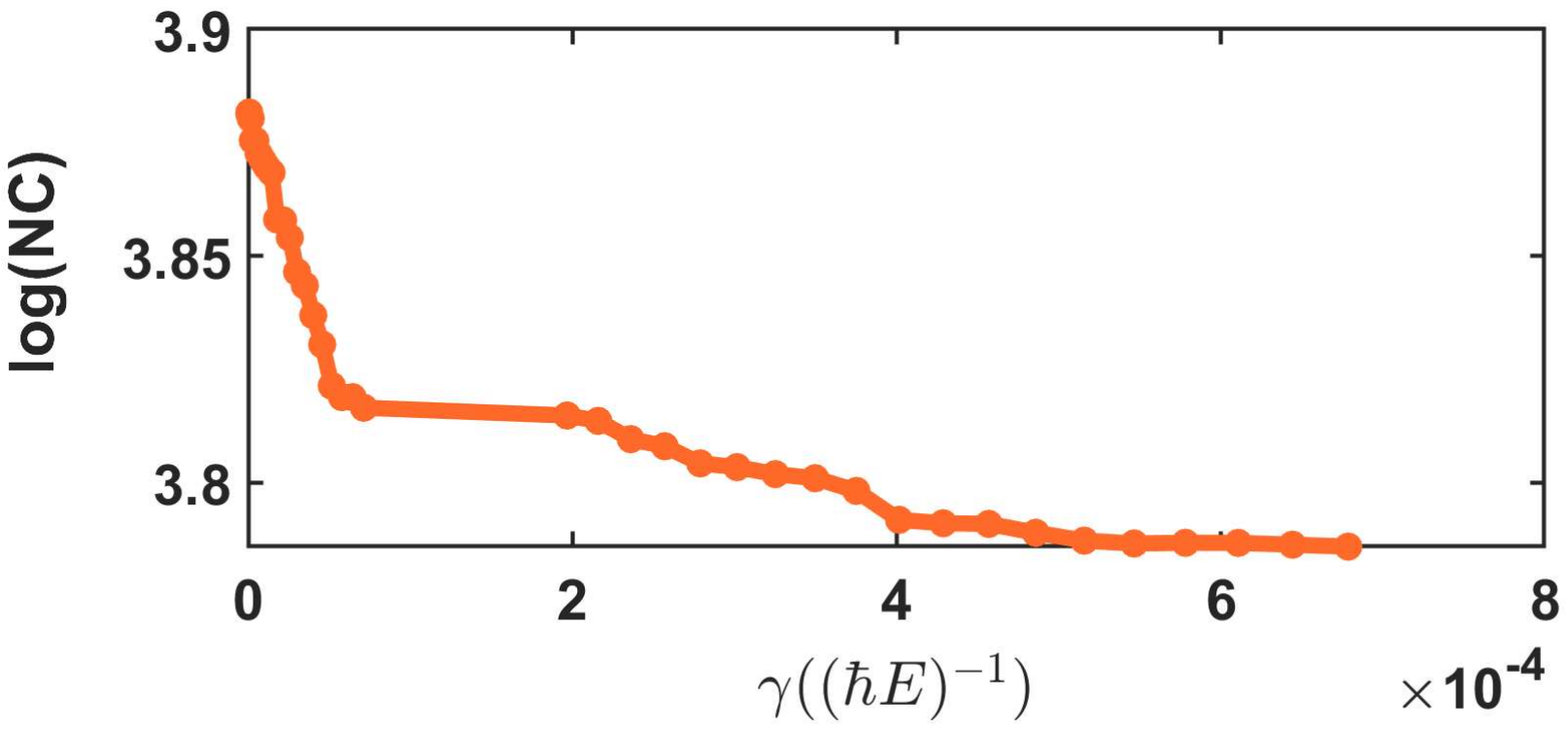}
    \vspace*{-10mm}\caption{\textbf{The noise cancellation $NC=\frac{IF_n}{IF_F}$ (log scale) Vs. the noise rate $\gamma$ for the entangling gate under amplitude noise.}}  
    \label{fig:Delta-G-H1-SU4-CNOT}
\end{figure}

The infidelity changes due to phase noise for the entangling SU(4) gate are shown in Fig. \ref{fig:Delta-G-H-SU4-CNOT}. 
The noise significantly degrades the fidelity. Optimal control can restore high fidelity. When the noise rate is low, we overcome the saturation 
in the Krotov method and obtain
infidelities better than the unitary case
(Cf. the inset), improving with noise rate. 
The reason is that Krotov's iterative procedure for systems with noise stagnates much later. New control channels are introduced to the control protocol, as suggested in \cite{wu2007controllability}.
For higher noise values, the mitigation due to OCT declines as expected.
The effect of amplitude noise on infidelity is shown in 
Fig. \ref{fig:Delta-G-H1-SU4-CNOT}. 
Optimal control managed to restore $\sim 4$ orders of magnitude. As expected, this mitigation decreases with the growth of the dephasing rate.

Additional insight is obtained using the optimal control field with noise for noiseless propagation.
The resulting infidelity is $ IF_U \sim 10^{-3}$, 3 orders of magnitude lower than the optimized $ IF_U \sim 10^{-6}$. This means that the optimal field with the noise represents a new active solution that emerges due to the control under noise.

\section{Discussion}
\label{sec:discussion}

A theoretical analysis can rationalize the fidelity loss of the gates due to the noise. Two factors can cause the loss: 
The first is a misguided unitary without loss of purity.
On the Bloch sphere, it reaches an erroneous final position. Such errors can be correctable with the aid of an additional coherent tool. The second element is an unrecoverable loss of purity. On the Bloch sphere, motion toward the interior of the sphere.

\trr{For a quantum state $\hat{\boldsymbol{\rho}}$ the} instantaneous purity loss is calculated as follows:
\begin{equation}
\frac{d}{dt} \textrm{tr} \{ \hat{\boldsymbol{\rho}}^2 \}~=~2 \textrm{tr} \{\hat{\boldsymbol{\rho}}  \boldsymbol{ \mathcal L} \hat{\boldsymbol{\rho}}\} 
~=~2\textrm{tr} \{\hat{\boldsymbol{\rho}}  \boldsymbol{ \mathcal D} \hat{\boldsymbol{\rho}}\} ~~,
  \label{eq:purityloss}
\end{equation}
The second equality results
from the conservation of purity by unitary transformations.
For state-to-state control for pure dephasing, the optimal solution is adiabatic, following \cite{levy2018noise}, where $\boldsymbol{ \mathcal D}_P \hat{\boldsymbol{\rho}}=0$. Nevertheless,
due to second-order terms in the Magnus approximation, the degradation due to phase noise does not vanish even for infinitely long control duration \cite{feldmann2010minimal}.

Phase noise is equivalent to the noise generated 
by imperfect timekeeping, where the fidelity has been found to scale as \cite{xuereb2023impact}:
$F=({2+e^{-\frac{\theta^2}{2 N}}})/{3}$,
where $\theta$ is the pulse area
and $N \propto 1/\gamma$ is the ticking accuracy. When calculating the fidelity as a function of $\gamma$, our results align with this scaling (Cf. Fig. \ref{fig:time-keeping}).

For amplitude noise, the instantaneous 
purity loss becomes in vector notation:
\begin{equation}
\begin{aligned}
\label{eq:purity}
    \frac{d}{dt} \textrm{tr} \{ \overrightarrow{\rho}^2 \}=2 \left(\overrightarrow{\rho} \cdot \tilde{\boldsymbol{\mathcal{ D}}}_A \overrightarrow{\rho} \right ) =  \\ -2{\gamma}_A\varepsilon_c^{2}(t)\left (\overrightarrow{\rho} \cdot {\boldsymbol{\mathcal H}}'_c{\boldsymbol{\mathcal H}}'_c \overrightarrow{\rho} \right)
    \end{aligned}
\end{equation}
For a pure state, the rate of purity loss for the amplitude noise model becomes \cite{boixo2007generalized}:
\begin{equation}
\frac{d}{dt} \textrm{tr} \{ \hat{\boldsymbol{\rho}}^2 \}~=~ -2 \gamma \varepsilon^2(t) \Delta {\boldsymbol{\mathcal H}}'_c
  \label{eq:purityloss-p}
\end{equation}
where $\Delta {\boldsymbol{\mathcal H}}'_c$
is the variance of the control operator \cite{khasin2011noise}.
Examining Eq. (\ref{eq:purityloss-p}) and considering that the pulse area is linear in $\varepsilon$
and the noise is quadratic,
the optimal solutions will minimize the control amplitude as the noise rate increases (Cf. Fig. \ref{fig:Delta-G-2-Had} and Fig. \ref{fig:Delta-G-6-Had}) or employ longer periods.

\trr{In a state-to-state control, the purity loss can be minimized by following a path that minimizes the variance with respect to the noise operator
$\left (\overrightarrow{\rho} \cdot {\boldsymbol{\mathcal H}}'_c {\boldsymbol{\mathcal H}}'_c \overrightarrow{\rho} \right)$ in Eq. (\ref{eq:purity}). 
For quantum gates, it is impossible to sito minimize a complete
base of states during propagation simultaneously therefore, observe
that for the optimal solution, this term is almost constant
when averaged over a complete set of initial states.}

Two general strategies of control mitigating the noise can be envisioned. The first is selecting the unitary control that minimizes the noise
from all possible controls.
The second is to employ the noise to modify the control trajectory. For thermal noise, we found the first mechanism to be operative \cite{kallush2022controlling}.
For controller noise, we observe that
the control solution fails when 
applied to noiseless dynamics.
As observed in Fig. \ref{fig:bloch}, the optimal solution with noise finds a different trajectory to the target.

An alternative  global mitigation idea is to post-process the gate operation by applying the inverse gate \cite{henao2023adaptive}
and combining it with the forward one. This method can be employed in the present context by doubling the control time.

\section{Conclusions}
\label{sec:conclus}

Quantum devices employ interference and entanglement as crucial resources. Dissipation is the primary limiting factor to achieving quantum technology \cite{haffner2008quantu,preskill2018quantum,schlosshauer2019quantum}. Significant effort is therefore devoted to isolating the quantum device from the
environment. However, the noise
originating from the controller is typically ignored. 
Since such noise is unavoidable, we address the question: Can optimal theory achieve the desired quantum gate while mitigating the harm due to noise? 

What is unique about controller noise is that upon "hands-off," the damage is avoided.
Our Markovian noise model assumes the controller is fast. A GKLS master equation describes the quantum dynamics of both amplitude and phase noise. Extension of this study to non-Markovian noise model \cite{kiely2021exact} and
Poissonian noise is feasible.

Employing advanced techniques of OCT
for open systems, we first find a control that generates a high-fidelity unitary gate. We then use this as a reference gate to 
study the degradation of fidelity due to the controller noise. 
Typically, the system is more sensitive to phase noise.
We then apply OCT again to mitigate 
the fidelity loss due to noise.
We have found that quantum control can find control fields that restore or surpass noiseless fidelity. \trr{Since the noise is sensitive to field intensity, a mechanism for reducing control noise is the reduction of field energy. Other mechanisms are more intricate.}

The present study would not have been possible 
without a highly accurate propagator addressing explicit time dependence
and non-hermitian operators (see Appendix). The gate infidelity is extremely small
but significant, and the differences between the unitary and noisy solutions are minute \cite{fernandes2023effectiveness}.

Our findings emphasize the significance of error mitigation techniques tailored to specific quantum gates and system characteristics. The varying quality of fidelity improvement based on the target gate and system underscores the need for targeted and adaptable strategies in the face of noise \cite{PhysRevApplied.20.024034}.

Notably, the integration of OCT into the framework of open quantum systems has led us to discover novel control solutions distinct from noiseless optimal solutions. In some cases, these control fields are able to restore more than three orders of magnitude in infidelity.

In conclusion, our study showcases the potential of OCT in mitigating errors caused by phase and amplitude noise, elevating gate fidelity closer to a feasible working framework within the domain of quantum computing.

\section*{Acknowledgment}
We thank Ido Schaefer 
for your help in implementing the semi-global propagator. In addition
we thank Raam Uzdin and Roie Dann 
for helpful discussions.
The Israel Science Foundation supported this study (grant nos. 510/17 and 526/21).

\appendix
\section{Vectorizing Liouville space}
\label{appndix:vector}

Propagators or solvers of dynamical equations of motion approximate the solution in a polynomial series \cite{kosloff1994propagation}. 
The basic primary entry step is 
matrix-vector multiplication. 
To employ such
propagators for the current open system dynamics in Liouville space, the numerical scheme has to be adopted. 

The following appendix describes the execution of superoperators acting on operators. This requires vectorizing Liouville space to adopt the standard matrix-vector operation. 
A description of the vectorizing procedure is employed in both analytical and numerical descriptions.

A Hilbert space composed of operators can 
be generated by defining a scalar product between operators. This is equivalent to a linear space of matrices, converting the matrices effectively into vectors $(\rho \rightarrow|\rho\rangle\rangle)$. This is the Fock-Liouville space (FLS) \cite{manzano2020short}. The usual definition of the scalar product of matrices $\phi$ and $\rho$ is defined as $\langle\langle\phi \mid \rho\rangle\rangle \equiv \operatorname{Tr}\left[\phi^{\dagger} \rho\right]$. The Liouville superoperator from Eq. (\ref{eq: von Neumann equation}) is now an operator acting on the Hilbert space composed of operators. The main utility of the FLS is to allow the matrix representation of the evolution operator.
For example, the evolution of a two-level system density matrix can be expressed in the FLS as
\begin{equation}
   |\rho\rangle\rangle=\left(\begin{array}{c}
\rho_{00} \\
\rho_{01} \\
\rho_{10} \\
\rho_{11}
\end{array}\right) . 
\end{equation}
The Liouvolle von Neumann equation describes the time evolution of a mixed state Eq. (\ref{eq: von Neumann equation}). In vector notation, the Liouvillian superoperator is expressed as a matrix,
\begin{equation}
   \tilde{\boldsymbol{\mathcal{L}}}=\left(\begin{array}{cccc}
0 & i \Omega & -i \Omega & 0 \\
i \Omega & i E & 0 & -i \Omega \\
-i \Omega & 0 & -i E & i \Omega \\
0 & -i \Omega & i \Omega & 0
\end{array}\right), 
\end{equation}
Each row is calculated by observing the output of the operation $-i[H, \rho]$ in the computational basis of the density matrices space. The system's time evolution corresponds to the matrix equation $\left.\frac{d|\rho\rangle\rangle}{d t}=\tilde{\mathcal{L}}|\rho\rangle\right\rangle$, which in matrix notation would be
\vspace{0.5cm}
\begin{equation}
    \left(\begin{array}{l}
\dot{\rho}_{00} \\
\dot{\rho}_{01} \\
\dot{\rho}_{10} \\
\dot{\rho}_{11}
\end{array}\right)=\left(\begin{array}{cccc}
0 & i \Omega & -i \Omega & 0 \\
i \Omega & i E & 0 & -i \Omega \\
-i \Omega & 0 & -i E & i \Omega \\
0 & -i \Omega & i \Omega & 0
\end{array}\right)\left(\begin{array}{l}
\rho_{00} \\
\rho_{01} \\
\rho_{10} \\
\rho_{11}
\end{array}\right) .
\end{equation}
A similar approach is used for the dissipative part $\tilde{\boldsymbol{\mathcal{D}}}$.

\section{The Semi-global Propagation Method}
\label{appendix:prop}

To solve the Liouville von Neumann equation, obtaining high-fidelity solutions for controlling quantum gates requires 
highly accurate and efficient numerical propagators to solve the Liouville von Neumann equation.
For this task, we have modified the semi-global propagator \cite{schaefer2017semi} to perform in
the Liouville vector space.

The Liouville von Neumann equation describes the dynamics of an open quantum system:
\begin{equation}
\frac{d}{d t} \hat{\boldsymbol{\rho}}={\cal L}\hat{\boldsymbol{\rho}}
\end{equation}
where $\hat{\boldsymbol{\rho}}$ is the density operator and ${\cal L}$ is the Liouvillian.
For a driven open system, the Liuouvillian can be partitioned to:
\begin{equation}
\begin{aligned}
\frac{d}{d t} \hat{\boldsymbol{\rho}}={\cal L}_t \hat{\boldsymbol{\rho}} =\left(
{\cal L}_H(t)+{\cal L}_D(t)\right)\hat{\boldsymbol{\rho}}\\ 
{\cal L}_H={\cal L}_{H_0}+{\cal L}_{H_t}
\end{aligned}
\label{eq:tlio}
\end{equation}
$\mathcal{L}_H(t)$ is the generator of the unitary part of the dynamics and can be decomposed into time-independent and time-dependent components. 
The dissipative part $\mathcal{L}_D(t)$ implicitly describes the effect of the environment and can be time-dependent. 

For a time-independent Lindblidian ${\cal L}_0$, a formal solution becomes:
\begin{equation}
    \hat{\boldsymbol{\rho}}{(t)}=e^{{\cal L}_0 t}\hat{\boldsymbol{\rho}}(0)
    \label{time-independent Lindblidian}
\end{equation}
with the initial conditions $\hat{\boldsymbol{\rho}}(0)$.
\\
\\
\\
\\
When the Liouviilan can be partitioned into a time-dependent and time-independent part
${\mathcal L}={\mathcal L}_0+{\mathcal L}_t$,
a formal solution of Eq. (\ref{eq:tlio}) can be written as an integral equation:
\begin{equation}
\hat{\boldsymbol{\rho}}(t)=e^{{\cal L}_0 t} \hat{\boldsymbol{\rho}}(0)+ \int_0^t e^{{\cal L}_0 (t-\tau)}{\cal{L}}_t \hat{\boldsymbol{\rho}}(\tau) d \tau
\label{integral-sulotion}
\end{equation}
Eq. (\ref{integral-sulotion}) will form the basis for the numerical approximation. 

In typical control problems 
$\mathcal L$  varies considerably with time.
Therefore, the total evolution is practically broken into finite time steps, $\Delta t$.
Then, one can concatenate the propagators and obtain the total evolution from $t=0$ to $t=T$ by
\begin{equation}
\overrightarrow{\mathbf{\rho}}(T) \approx \prod_{j=1}^{N_t} {\mathcal G }_j(\Delta t) \overrightarrow{\mathbf{\rho}}(0)
\label{Prudoct-rule}
\end{equation}
where ${\mathcal G}_j(\Delta t)$ is the 
propagator for $t$ to $t+\Delta t$ and $t=j \Delta t$. 

A direct approximation assumes that ${\mathcal L}_t$ is time independent within a time step, then
\begin{equation}
    {\mathcal G_j} \approx e^{ {\mathcal L}_j \Delta t}
    \label{eq:gprop}
\end{equation}
where ${\mathcal L}_j = {\mathcal L}(t+\Delta t/2)$. Sampling ${\mathcal L}$
in the middle of the time step leads to second-order accuracy in $\Delta t$.

Subsequently, we formulate the density operator as a vector $\overrightarrow{\mathbf{\rho}}$
and the Liouvillian superoperator ${\cal L}$
as a matrix ( see above App. A \cite{am2015three}).

A numerical method to solve  Eq.(\ref{eq:gprop}) is based on expanding the exponent or any analytic function in a polynomial series in the matrix ${\cal L}_j$:
\begin{equation}
 \overrightarrow{\mathbf{\rho}}(t+\Delta t) \approx \sum_{n=0}^{K-1} a_n(t+\Delta t) P_n\left(\mathcal{L}_j\right) \overrightarrow{\mathbf{\rho}}(t) 
 \label{Polynomial expention}
\end{equation}
where $P_n(\mathcal{L}_j)$ is a polynomial of degree $n$, and $a_n(t+\Delta t)$ is the corresponding expansion coefficient in the interval $t,t+\Delta t$. This requires choosing the set of expansion polynomials $P_n(\mathcal{L}_j)$ and the corresponding coefficients $a_n$  \cite{berman1991time}. 

The expansion (\ref{Polynomial expention}) has to be accurate in the eigenvalue domain of $\mathcal{L}_j$ so that the form (\ref{Polynomial expention}) will be helpful for the representation of $ {\mathcal G_j} $.
Successive matrix-vector multiplications can compute this series of polynomials at Eq. (\ref{Polynomial expention}). It scales as \trr{$O\left(KN^2\right)$}, and the computational effort can be reduced even further. For sparse superoperators, the matrix-vector operation can be replaced by an equivalent semi-linear scaling with \trr{$\sim KO(N)$} \cite{kosloff1983fourier}. 

An immediate question is how to choose the set of expansion polynomials $P_n(x)$. As a thumb rule, we search for a polynomial basis that converges the fastest. An orthogonal set of polynomials is the first step for fast convergence \cite{kosloff1994propagation}.

An efficient implementation can be done recursively. The Chebyshev polynomial and Newton interpolation polynomials are two orthogonal expansion series using $P_n(\mathcal{L}_j)$. 
When the Hamiltonian is non-Hermitian, the eigenvalue domain becomes complex, and the Chebychev approach is not appropriate anymore. Then, the Newtonian or Arnoldi approach should be used instead \cite{kosloff1994propagation,arnoldi1951principle,lehoucq1996deflation}. 

Note that in Eq. (\ref{time-independent Lindblidian}), 
only the coefficients $a_n(t)$
are time-dependent.
The solution  at intermediate time points can be obtained by calculating the coefficients
for intermediate points
with negligible additional computational effort.

Quantum control of gates requires exceptionally high accuracy. The convergence rate of Eq. (\ref{Prudoct-rule}) with a piecewise constant ${\mathcal L}_j$ is slow, leading to extensive numerical effort. To obtain faster convergence, we must consider the problem of time ordering within the time step $\Delta t$.
To overcome the problem of time ordering, we will combine the polynomial solution of Eq. (\ref{time-independent Lindblidian}) and the integral equation formal solution (\ref{integral-sulotion}).
In Eq. (\ref{integral-sulotion}), the free propagator appears both as a free term and at the integrand. The solution of the integral equation requires an iterative approach since 
$\hat{\boldsymbol{\rho}}(\tau)$ also appears in the integrand.
The iteration is carried out by extrapolating the solution
from one time step to the next, $t$ to $t+dt$.
The integral in the formal solution Eq. (\ref{integral-sulotion}) is reformulated  
as an inhomogeneous source term:
\begin{equation}
    \frac{d \overrightarrow{\mathbf{\rho}}(t)}{d t}=\mathcal{L}_j \overrightarrow{\mathbf{\rho}}(t)+\overrightarrow{\mathbf{s}}(t)
     \label{source-term}
\end{equation}
The source term will represent the time-dependent/nonlinear part of the dynamics.
Treating Eq. (\ref{source-term}) as an inhomogeneous ODE will give rise to a formal solution to the time-dependent problem.

We can write the solution to Eq. (\ref{time-independent Lindblidian}):
\begin{equation}
    \begin{aligned}
\overrightarrow{\mathbf{\rho}}(t+\Delta t) & =\tilde{ \boldsymbol{ \mathcal G}}_j(t) \overrightarrow{\mathbf{\rho}}(t) +\int_t^{t+\Delta t }\tilde{ \boldsymbol{ \mathcal G}}_j(t-\tau) \overrightarrow{\mathbf{s}}(\tau) d \tau \\
& =\exp \left(\mathcal{L}_j t\right) \overrightarrow{\mathbf{\rho}}+
\\
& \int_t^{t+\Delta t} \exp \left[\mathcal{L}_j(t-\tau)\right] \overrightarrow{\mathbf{s}}(\tau) d \tau \\
& =\exp \left(\mathcal{L}_j t\right) \overrightarrow{\mathbf{\rho}}_0 
+\\ &\exp \left(\mathcal{L}_j t\right) \int_t^{t+\Delta t} \exp \left(-\mathcal{L}_j \tau\right) \overrightarrow{\mathbf{s}}(\tau) d \tau&
\end{aligned}
\label{Dunhamel-principle}
\end{equation}

Where $\tilde{ \boldsymbol{ \mathcal G}}_j$ is defined as the time-independent propagator by the vec-ing procedure: subsection \ref{subsec:veccing}.

The source term is expanded as a time-dependent polynomial to solve for the integral analytically.
\begin{equation}
    \overrightarrow{\mathbf{s}}(t)=\sum_{n=0}^{M-1} \frac{t^n}{n !} \overrightarrow{\mathbf{s}}_n
    \label{source-term-polynomial}
\end{equation}
This expansion allows us to solve formally
the integral  in Eq. (\ref{Dunhamel-principle})
\begin{align*}
  \int e^{a t} t^m/m ! = \sum_{n=1}^m  e^{a t} t^{n-m}/a^n(n-m)!.  
\end{align*}

The problem is now shifted to obtaining the expansion coefficients $\overrightarrow{\mathbf{s}}_n$.
This task is obtained by
approximating $\overrightarrow{\boldsymbol{s}}(t)$ by an orthogonal polynomial in the time interval.
We choose a Chebychev expansion
\begin{equation}
    \overrightarrow{\mathbf{s}}(t) \approx \sum_{n=0}^{M-1} \overrightarrow{\mathbf{b}}_n T_n(t)
\label{eq:monomial}
\end{equation}
where the coefficients $\overrightarrow{\mathbf{b}}_n$ are computed by a scalar product of the $T_n(t)$ with  $\overrightarrow{\mathbf{s}}(t)$. 
Approximating the coefficients using Chebychev 
sampling points in the time interval $\Delta t$.

The coefficients $\overrightarrow{\mathbf{s}}_n$
are calculated by relating the polynomial Eq. (\ref{eq:monomial}) 
to the Chebychev expansion.
This source term is inserted into the integral  Eq. (\ref{Dunhamel-principle}), leading to a numerical approximation to the solution of the TDLE.
The addition of the source term into the dynamics gives rise to an analytical solution for the last term in Eq. (\ref{Dunhamel-principle}), presented here on the RHS of Eq.  (\ref{Recortion-integral})
\begin{equation}
    J_{m+1}(\mathcal{L}_j, t) \equiv \int_t^{t+\Delta t} \exp (-\mathcal{L}_j \tau) \tau^m d \tau, \quad m=0,1, \ldots
    \label{Recortion-integral}
\end{equation}
with the recursion relations:
\begin{equation}
\begin{aligned}
      J_m(\mathcal{L}_j, t)=-\frac{\exp (-\mathcal{L}_j t) t^{m-1}}{\mathcal{L}_j}+\frac{m-1}{\mathcal{L}_j} J_{m-1}(\mathcal{L}_j, t), \\ \quad m=2,3, \ldots  
\end{aligned}
\end{equation}
where
\begin{align}
    J_1(\mathcal{L}_j, t) \equiv \int_t^{t+\Delta t} \exp (-\mathcal{L}_j \tau) d \tau=\frac{1-\exp (-\mathcal{L}_j t)}{\mathcal{L}_j}.
\end{align}
Plugging Eq. (\ref{source-term-polynomial}) into this formulation leads to the following:
\begin{equation}
\begin{aligned}
     \exp (\mathcal{L}_j,t) \sum_{n=0}^{M-1} \frac{1}{n !} \int_0^t \exp (-\mathcal{L}_j \tau) t^n d \tau s_n= \\ \exp (\mathcal{L}_j t) \sum_{n=0}^{M-1} \frac{1}{n !} J_{n+1}(\mathcal{L}_j, t) s_n=\sum_{n=0}^{M-1} f_{n+1}(\mathcal{L}_j, t) s_n   
\end{aligned}
\end{equation}

In Eq. (\ref{eq:tlio}), the Louivilian is split into explicit time-dependent and approximated time-independent parts; this partition is used  to re-write the TDLE
\begin{equation}
   \frac{d \overrightarrow{\mathbf{\rho}}(t)}{d t}={\mathcal{L}}_j \overrightarrow{\mathbf{\rho}}(t)+\overrightarrow{\mathbf{s}}(\overrightarrow{\mathbf{\rho}}(t), t) 
\end{equation}
Which is a reformulation of Eq. (\ref{source-term}). The same analysis leads to:
\begin{equation}
\begin{aligned}
 \overrightarrow{\mathbf{\rho}}(t,t+\Delta t)=\exp (\mathcal{L}_jt) \overrightarrow{\mathbf{\rho}}(t)+ \\ \exp ({\mathcal{L}_j} t) \int_t^{t+\Delta t} \exp (-{\mathcal{L}_j} \tau) \overrightarrow{\mathbf{s}}(\overrightarrow{\mathbf{\rho}}(\tau), \tau) d \tau   
\end{aligned}
\label{eq:integ}
\end{equation}
Now, we can use these formulations to approximate Eq. (\ref{eq:integ}):
\begin{equation}
\begin{aligned}
    \overrightarrow{\mathbf{\rho}}(t,t+ \Delta t) \approx P_M({\mathcal{L}_j}, (t,t+ \Delta t)) \overrightarrow{S}(t,t+ \Delta t)_M+ \\ \sum_{n=0}^{M-1} \frac{t^n}{n !} \overrightarrow S(t,t+ \Delta t)_n  
\end{aligned}
 \label{Final-TDLE inhomogeneous linear ODE}
\end{equation}
 $ P_M({\mathcal{L}_j}, (t,t+ \Delta t) \overrightarrow{S} (t,t+ \Delta t)_M$, is approximated by the Arnoldi method (the eigenvalue spectrum of $\mathcal{L}$ is distributed on the complex plane) 
where 
\begin{align*}
            \overrightarrow{{S(t)}}_M \equiv  \overrightarrow{\mathbf{s}}(t)+\mathcal{L}_t \overrightarrow{\mathbf{\rho}(t)}
\end{align*}
 are computed by expanding it by time in the same form of (\ref{source-term-polynomial}). 
We have used here the fact that $P_n(\mathcal{L_{H}},t)=\mathcal{L_{H}}^{k-n} P_k(\mathcal{L_{H}}, t)+\sum_{j=n}^{k-1} \frac{t^j}{j !} \mathcal{L_{H}}^{j-n}$
Eq. (\ref{Final-TDLE inhomogeneous linear ODE}) and $\overrightarrow{\mathbf{s}}_j$ include dependence on $\overrightarrow{\mathbf{\rho}}(t)$. 
It would seem we are back to the same problem. However, it can be conquered through a process of repetition and refinement. 

First, we guess a solution $\overrightarrow{\mathbf{\rho}}_g(t)$, 
within a time step $\Delta t$, and use it in Eq. (\ref{Final-TDLE inhomogeneous linear ODE}) to obtain a new approximate solution. This procedure can be continued until the solution converges with the desired accuracy. The initial guess is extrapolated from the previous time step to accelerate convergence.

Three numerical parameters determine the precision of the propagation and the convergence rate:
\begin{itemize}
\item{The size of the time step $\Delta t$.}
\item{Number of Chebyshev sampling points in each time step $M$.}
\item{The size of the Krylov space $K$ corresponds to the basis of the Arnoldi algorithm. It is important to note that this parameter is limited by $Dim\{{\mathcal{L}\}-1}$}
\end{itemize}
Each of those is adjustable by the user to fit their needs best. For example, in the Hadamard transformation system, we use the following parameters: $\Delta t = 0.1$, $M = 7$, and $K=3$. With these parameters, we got an accuracy of $10^{-8}$ for the propagator, two orders of magnitude higher than the fidelity of the target transformation. For the entangling gate, we adjust the parameters for a higher resolution that would fit $10^{-8}$ for the fidelity of the target transformation ($M=K=9$).
\section{The Impact of Imperfect Time-keeping on Quantum Control}
\label{appndix:Timekeeping}
\begin{center}
   \vspace{-2.2cm}
    \includegraphics[width=0.95\linewidth,angle=-90]{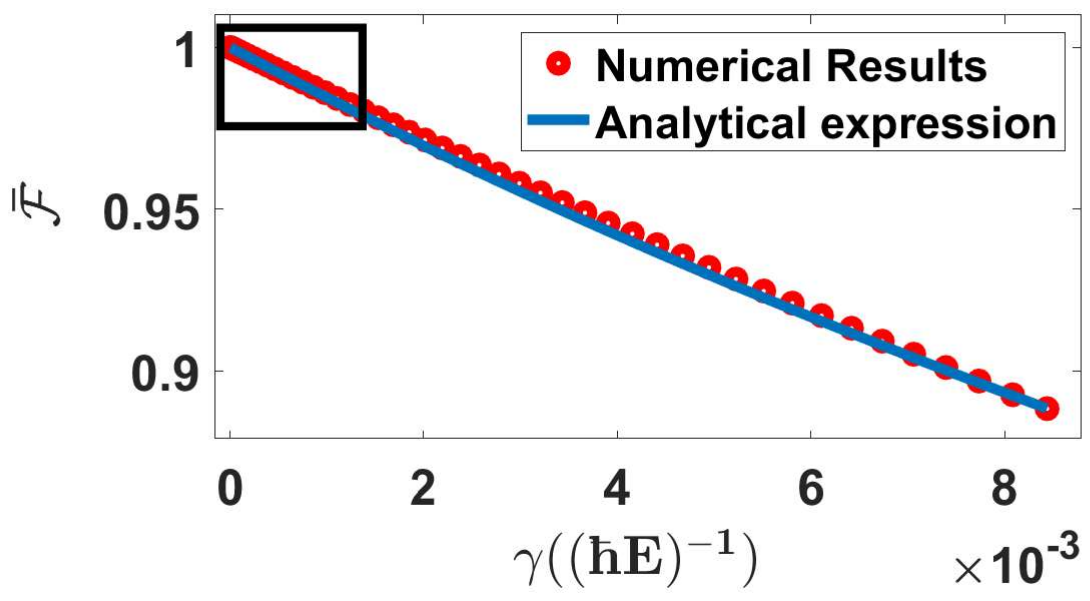}
        \vspace{-3.5cm}
        \captionof{figure}{The fidelity as a function of the noise rate
        for the Hadamard gate on a large scale. The dots are from our numerical calculations, and the line is 
        derived from the analytical expression from Ref. \cite{xuereb2023impact}.
        The square refers to the fidelity region considered in the mitigation by the optimal control.}
    \label{fig:time-keeping}
\end{center}

\section{Derivation of the correction to the control field}

The functional derivative with respect to the control field    $\varepsilon$, of the objective functional Eq. (\ref{eq:functional})\\ $\frac{d\mathcal{J}_{Tot}}{d\epsilon}=0$, leads to the correction to the field.

The update term for iteration $k$ in the Krotov method is defined as \cite{palao2003optimal}
$$
\Delta \epsilon_l^{(k)}(t) \equiv \epsilon_l^{(k)}(t)-\epsilon_l^{(k-1)}(t),
$$
where $\epsilon_l^{(k-1)}(t)$ is the previous iteration.
The penalty functional is now redefined by
\begin{equation}
   \mathcal{J}_{penal} =\lambda \int_0^T \frac{1}{s(t)}|\Delta \epsilon_l^{(k)}(t)|^2 d t
   \label{eq:limpower}
\end{equation}.
Now the derivative of each with respect to $\epsilon$ becomes:
\begin{equation}
    2\lambda \int_0^T \frac{1}{s(t)}|\Delta \epsilon_l^{(k)}(t)| dt
\end{equation}
The functional derivative of the dynamical constraint leads to (for amplitude noise):
\begin{equation}
        \int_0^T \operatorname{Tr}\left\{ \Lambda(t)^{k-1}\left(- {{\boldsymbol{\mathcal H}}'_c +2\Delta \epsilon {\boldsymbol{\mathcal H}}'_c {\boldsymbol{\mathcal H}}'_c} \Lambda(t)\right) \Upsilon(t)^k\right\} dt \\
\end{equation}
The two terms should sum to 0, which leads to the update equation for the field Eq. \eqref{eq:update}.
A similar approach can be used for phase noise, leading to the update equation:
\begin{equation}
    \begin{aligned}
          \Delta \boldsymbol{\epsilon}(t)=-{s(t)}\operatorname{Im}\frac{\text Tr\{\tilde{\boldsymbol Y}^*
          (\tilde {\boldsymbol{\mathcal H}}'_c +\{{\boldsymbol{\mathcal H}}'_c, {\boldsymbol{\mathcal H}}'_0\} )\tilde{\boldsymbol{ \mathcal G}}(t)  \}}{2( \lambda+ {\gamma}_A Tr\{\tilde{\boldsymbol Y}^* \tilde{\boldsymbol{ \mathcal H}}_c^2 \tilde{\boldsymbol{ \mathcal G}}(t)\})}  
    \end{aligned}
    \label{eq:update-phase}
\end{equation}

\section{Deriving Field} \label{Appendix E}
The driving control field $\varepsilon$ for specific representative cases of the Hadamard gate is displayed:
\begin{figure}[h!]
    \centering
    \includegraphics[width=1.\linewidth]{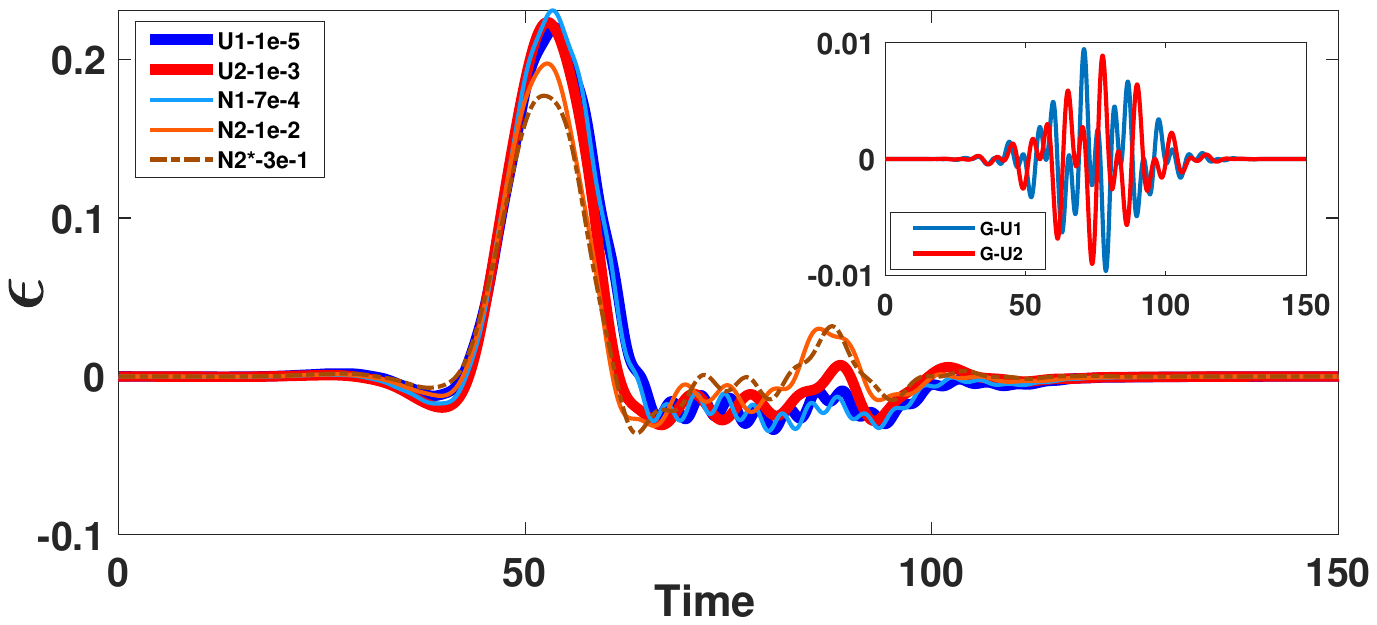}
    \vspace{-3.cm}
    \caption{The control field $\varepsilon$ as a function of time for the Hadamrad gate. This solution is initiated from two different guess fields presented in the insert. The first guess (U1) converges in the noiseless case to the infidelity of $IF=10^{-5}$, while the second guess (U2) converges to $IF=10^{-3}$.
    The converged fields are illustrated in the main graph as blue and red lines. This field is used as an initial guess field for the cases with noise. N1 (light blue)-corresponds to the guess used in U1, and N2(orange)-corresponds to the guess used in U2. N2*(dark red) shows the control field for a noise increase by a factor of five. }
    \label{fig:2-Guessed-solution}
\end{figure}
In Fig. \ref{fig:2-Guessed-solution}, we show two solutions obtained from two different initial guess fields using OC for the Hadamard gate. The solution for the first guess, U1, is the result of a system with no noise; the infidelity we got there is $1 \times 10^{-5}$, and the corresponding simulation with noise(N1), with the infidelity of $7 \times 10^{-4}$. The two fields are quite similar, with slight differences in later times. For the second example, U2, N2, and N2*, one can observe significant changes, especially in later times. These two examples highlight using a pilot field to accelerate the convergence of the optimal control solutions.


\end{document}